\documentclass[preprint,prd,aps,nofootinbib]{revtex4}
\usepackage{graphicx}
\usepackage{diagbox}
\usepackage{float}
\usepackage{mathrsfs}
\usepackage{color}
\usepackage{slashed}
\begin{document}
\title{Identification of $D^*_2(3000)$ as the $D_2^*(2^3P_2)$ and exploring potential of undiscovered $2^+$ mesons via $B$ decays}
\author{Shi-Hang Zhang$^{1,2,3,*}$,
 Wen-Yuan Ke$^{1,2,3}$\footnote{These authors contributed equally},
 Su-Yan Pei$^{1,2,3}$,
 Wei Li$^{4}$,
 Xiao-Ze Tan$^{5,6}$,
 Lili Zhu$^{7}$,
Guo-Li Wang$^{1,2,3}$\footnote{Corresponding author}}
\affiliation{$^1$ Department of Physics, Hebei University, Baoding 071002, China\\
$^2$ Hebei Key Laboratory of High-precision Computation and
 Application of Quantum Field Theory, Baoding 071002, China\\
$^3$ Hebei Research Center of the Basic Discipline for Computational Physics, Baoding 071002, China\\
$^4$ College of Science, Hebei Agriculture University, Baoding 071001, China\\
$^5$ Department of Physics and Center for Field Theory and Particle Physics, Fudan University, Shanghai 200438, China\\
$^6$ Deutsches Elektronen-Synchrotron DESY, Notkestr. 85, 22607 Hamburg, Germany\\
$^7$ College of Science, Beihua University, Jilin 132013, China}
\begin{abstract}
Following the discovery of the $D^*_2(3000)$, its mass and full width have been extensively studied. Yet its nature remains undetermined to date. Since it was discovered through nonleptonic decay of $B$ meson and the corresponding cascade process, we therefore
in this paper investigate the nonleptonic and  semileptonic decays of $B$ meson to $J^P = 2^+$ charmed mesons using the Bethe-Salpeter equation approach. Our calculations on nonleptonic $B$ decays reveal that the unconfirmed resonance $D^*_2(3000)$ aligns well with $D^*_2(2^3P_2)$ predictions. Other candidates, including $D^*_2(1^3F_2)$, $D^*_2(3^3P_2)$, and $D^*_2(2^3F_2)$, are excluded due to their very small branching ratios in $B$ decays. Considering that the $D^*_2(1F)$, $D^*_2(3P)$, and $D^*_2(2F)$ have not yet been experimentally observed, we investigate the feasibility of their detection in $B$-meson decays.
\end{abstract}
\maketitle
\section{Introduction}
In recent years, experimental studies have made significant progress, leading to the discovery of numerous new particles, including excited-state mesons and exotic hadronic states, resulting in a rapid expansion of the hadron spectrum.
However, certain theoretically predicted excited-state mesons, anticipated long ago, have still not been experimentally discovered, or candidate particles have been found but cannot be confirmed. For instance, several charmed mesons have been experimentally discovered near the region of 3000 MeV.
In 2013, the LHCb collaboration observed $D_J(3000)$ and $D^*_J(3000)$ \cite{1}. Later in 2016, LHCb further discovered $D^*_2(3000)$ \cite{2}.

Current analyses suggest that $D_J(3000)$ could correspond to a high-excitation state with $J^P=0^-$ pseudoscalar  $D(3^1S_0)$ \cite{cankaoshu3,lx1,ldm1} or a $1^+$ state $D_1(2P'_1)$ \cite{cankaoshu7,lsc1,lsc2}. The $D^*_J(3000)$ might represent a $1^+$ state $D_1(2P_1)$ \cite{batra,lx2}, a $0^+$ state $D^*_0(2^3P_0)$ \cite{cankaoshu8,txz2}, a $2^+$ state $D_2^*(1^3F_2)$ \cite{wzg1}, or a $4^+$ $D^*_4(1^3F_4)$ \cite{zxh1}. We note that for the $D_J(3000)$ and $D^*_J(3000)$, only their masses and widths are currently known in experiment, while their $J^P$ quantum numbers cannot be determined. Moreover, the LHCb collaboration has only mentioned their reconstruction in the $D^{*+}\pi^-$ ($D_J(3000)^0$), $D^{+}\pi^-$ ($D^*_J(3000)^0$) and $D^{0}\pi^+$ ($D^*_J(3000)^+$) final states, without providing details on their specific production processes or other information such as branching fractions. Therefore, the properties of these two particles cannot be determined based solely on the currently available experimental information.

As for $D^*_2(3000)$, its quantum number $J^P=2^+$ is confirmed, and its mass and width are
\begin{eqnarray}
&&M(D^{*0}_2(3000))=3214\pm29\pm33\pm36~\rm{MeV},\nonumber \\ &&\Gamma(D^{*0}_2(3000))=186\pm38\pm34\pm63~\rm{MeV},
\end{eqnarray}
with large uncertainties.
By calculating its mass and estimating its full width, it was found that it could be a $2^+$ tensor meson such as $D^*_2(2^3P_2)$ \cite{txz1,cankaoshu8,Fazio}, $D_2^*(1^3F_2)$ \cite{Fazio,wzg2,cankaoshu9,cankaoshu8},  $D^*_2(3^3P_2)$ \cite{7}, or $D_2^*(2^3F_2)$  \cite{7,txz1}. Overall, the nature of this particle also remains incompletely determined.

The challenges in determining the properties of these particles stem from the following factors: Experimentally, the mass measurements of these particles lack sufficient accuracy, and more critically, there are substantial uncertainties in their total decay widths.
Theoretically, research has predominantly focused on calculating their masses and total widths. As noted earlier, the 3000 MeV region hosts numerous candidate states, leading to significant ambiguities. Furthermore, theoretical calculations of total widths face major uncertainties due to:
1. The presence of multiple OZI-allowed strong decay channels for these particles. The partial widths of such decays above the thresholds are highly sensitive to mass values, amplifying theoretical uncertainties.
2. The existence of 1-2 nodes in the wave functions of excitation $2P$, $2F$, $3S$ and $3P$ states. Due to the sign reversal of the wave function values before and after the nodes, their contributions to the transition amplitude cancel out, consequently, theoretical predictions for decay width are critically dependent on the positions of these nodes, leading to significant errors arising from slight differences in mass and other parameters.
This interplay of experimental limitations and theoretical sensitivities complicates the precise characterization of these particles.

However, unlike the $D_J(3000)$ and $D^*_J(3000)$, we note that the experiment identified the $D^*_2(3000)$ through its reconstruction in the final states of non-leptonic $B$-meson decays and provided the product of branching ratios  \cite{2},
\begin{eqnarray}
&&\mathcal{B}\left(B^{-} \rightarrow \bar{D}_{2}^{*}(3000)^0 \pi^{-} \right) \mathcal{B}\left(\bar{D}_{2}^{*}(3000)^0 \rightarrow D^{+} \pi^{-}\right) =  (2 \pm 1\pm1\pm1) \times 10^{-6}.
\end{eqnarray}
Crucially, combined with mass and width of $D^*_2(3000)$, this product value of branching ratios opens up the possibility for ultimately determining the nature of the $D^*_2(3000)$.
Therefore, in this paper, we will focus on investigating the semileptonic decays of $B\rightarrow {D}_2^{*}(nP)\ell^+\nu_{\ell}$ and
$B\rightarrow D_2^{*}(mF)\ell^+\nu_{\ell}$, and the nonleptonic decays $B\rightarrow {D}_2^{*}(nP)\pi^+$ and ${B}\rightarrow D_2^{*}(mF)\pi^+$, where $n=1,2,3$ and $m=1,2$. We expect to first determine the properties of $D^*_2(3000)$, identifying which meson corresponds to $D^*_2(3000)$, and then explore the possibility of discovering other undetected particles in $B$-meson decays.

Beyond clarifying the properties of $D^*_2(3000)$, investigating the nature of $2^+$ states itself holds significant importance. This is because, apart from the ground $2^+$ meson $D^*_2(2460)$, our knowledge of excited states such as $2P$ and $3P$ with orbital angular momentum $L=1$ remains limited. Moreover, our understanding is even scarcer regarding $1F$ and $2F$ $2^+$ states (with $L=3$). Furthermore, analogous to the $S-D$ mixing state $\psi(3770)$, theory predicts the existence of $P-F$ mixing in $1F$ and $2F$ mesons, yet these have never been experimentally confirmed. Therefore, conducting research on $2^+$ states will enhance our comprehension of $P-F$ mixing phenomena.

The reason for studying the productions of $D^*_2(nP)$ and $D^*_2(mF)$ mesons in $B$ meson decays is that we currently have three global $B$-meson factories, Belle, BaBar, and LHCb, all of them have accumulated enormous numbers of $B$-meson events. These datasets not only facilitate investigations into $B$-meson semileptonic decays, nonleptonic decays \cite{lilun1,lilun2,lilun3,lilun5,shiyan6,shiyan7,shiyan8,shiyan9,shiyan10,shiyan11}, and Cabibbo-Kobayashi-Maskawa (CKM) matrix elements \cite{ckm1,ckm2}, but also enable studies of $CP$ violation  \cite{cp1,cp2}, searches for new physics \cite{xinwuli1,xinwuli2}, and more. The vast $B$-meson event statistics have made its decay processes a leading source for discovering new particles in recent years. Consequently, searching for highly excited $D^*_2$ mesons in the final states of $B$ decays represents the most competitive strategy.

As emphasized in Refs. \cite{19,wgl1,liwe}, relativistic corrections for excited heavy mesons cannot be neglected, and these corrections become increasingly significant for higher excited states. Consequently, studying the high-excitation state such as the $D^*_2(1F)$ and $D^*_2(3P)$  necessarily requires a relativistic method, as non-relativistic models would introduce substantial errors. In this work, we therefore adopt the Bethe-Salpeter (BS) equation method \cite{BS}. In quantum field theory, the relativistic dynamical equation for bound states is the BS equation, which serves as the foundation for all bound state theories. However, due to its complexity, it has not been possible to solve this equation exactly to this day, and approximations must be employed for its application. Salpeter pointed out that for heavy mesons, the instantaneous approximation is a valid approach, and by applying this approximation to the BS equation, he derived the Salpeter equation  \cite{Salpeter}. The Salpeter equation has been widely used \cite{Brodsky,Thompson,Lucha,Colangelo,Resag,Munz,ktchao,Loringa}, but researchers have not solved the full Salpeter equation (which consists of four coupled equations). Instead, they have only solved the most important one of these equations, thereby discarding a significant amount of relativistic information.

Over the past two decades, we have solved the complete Salpeter equations for various types of hadrons \cite{0-,1-,Pwave}. The obtained results were applied to studies of mass spectra \cite{spec,wgl2}, weak decay processes \cite{fuhf,19,zhout}, strong interaction processes \cite{ann0-,wgl5,3930}, and electromagnetic processes \cite{li01,pei02,li02,pei01}, etc, all yielding results in excellent agreement with experimental data, thereby validating the effectiveness of our methodology. In our approach, to formulate the expression of the relativistic wave function, we abandon the non-relativistic $^{2S+1}L_J$ notation and instead construct the wave function expression based on the $J^P$ of the particle, ensuring that each term in the wave function possesses the same $J^P$ as the particle itself. The undetermined radial wave functions are derived by solving the full Salpeter equation. In the resulting wave functions, the non-relativistic components and relativistic corrections each exhibit distinct Lorentz structures.
Notably, the wave function is not a pure wave as in non-relativistic approaches but inherently incorporates multiple partial waves.
This implies that we do not need to artificially mix $P$-wave and $F$-wave to achieve $P-F$ mixing, this mixing of distinct partial waves automatically emerges within the relativistic wave function. Furthermore, our results reveal not merely $P-F$ mixing, but a more comprehensive $P-D-F$ mixing \cite{wgl2}, where the relativistic framework naturally incorporates intertwined configurations of multiple partial waves.

The paper is organized as follows. {In Sec. II, the Bethe-Salpeter equation and the Salpeter equation are presented. In Sec. III, we will introduce the formulas for semileptonic and nonleptonic decays.
Sec. IV provides the relativistic wave functions of the mesons used in our calculations. We show the results and discussions in Sec. V.
Finally, Sec. VI summarizes the paper.}

{\section{The Bethe-Salpeter equation and the Salpeter equation}

The BS equation is a relativistic dynamical equation for bound states. For a meson composed of a quark and an antiquark, its representation is \cite{BS}:
\begin{eqnarray}
(\slashed{p}_2-m_2)\chi_{_{_P}}(q)(\slashed{p}_1+m_1)=i\int\frac{d^4k}{(2\pi)^4}V(P,k,q)\chi_{_{_P}}(k),
\end{eqnarray}
$p_1$ and $p_2$ denote the momenta of the antiquark and quark, respectively. $m_1$ and $m_2$ represent the masses of the antiquark and quark. $V(P, k, q)$ denotes the integral kernel between the quarks in the bound state. $\chi_{_{_P}}(q)$ is the BS wave function of the meson, where $q$ is the relative momentum between the quarks, and $P$ is momentum of the meson.
The relationship between the external momentum $P$ and the internal relative momentum $q$, can be expressed as: $p_1=\alpha_1P-q$,
$p_{2}=\alpha_2P+q$, where $\alpha_1=\frac{m_1}{m_1+m_2}$, $\alpha_2=\frac{m_2}{m_1+m_2}$.

Solving the four-dimensional Bethe-Salpeter equation is highly challenging. Since the instantaneous approximation is well-suited for heavy mesons, Salpeter applied it to the BS equation and  reduced it to the three-dimensional Salpeter equation. The core idea of this approximation is to neglect the interaction propagation time between the quarks inside the meson, meaning the integral kernel $V$ is independent of the time component or the zeroth component of the momentum. Thus, in the meson's center-of-mass system (CMS):
\begin{eqnarray}
V(P,k,q)\simeq V(\vec{k},\vec{q})=V(k_{_\bot},q_{_\bot}),
\end{eqnarray}
where $q_{_{\bot}}\equiv q-\frac{P\cdot q}{M}P$, and $q_{_{\bot}}=(0,\vec{q})$ in the CMS of $P$. In our calculations, the Cornell potential is employed, with its explicit expression provided in Appendix B.

Define the following two functions \cite{Salpeter}:
\begin{eqnarray}
&&\varphi_{_P}(q_{_\bot})\equiv i \int\frac{dq_{_{_P}}}{2\pi}\chi_{_{_P}}(q),\nonumber\\
&&\eta(q_{_\bot})\equiv\int\frac{d k_{_\bot}}{(2\pi)^3}V(k_{_\bot},q_{_\bot})\varphi_{_P}(k_{_\bot}),
\end{eqnarray}
where $\varphi_{_P}(q_{_\bot})$ denotes the Salpeter wave function. Thus, the BS equation becomes:
\begin{eqnarray}\label{bse}
&&\chi_{_P}(q)=S_2(p_2)\eta(q_{_\bot})S_1(p_1).
\end{eqnarray}
The propagator $S_i$ ($i=1$ for antiquark and $i=2$ for quark) can be decomposed as:
\begin{eqnarray}\label{propa}
S_i(p_i)=\frac{\Lambda^+_{iP}(q_{_\bot})}{J(i)q_{_{_P}}+\alpha_iM-\omega_{_{iP}}+i\epsilon}+\frac{\Lambda^-_{iP}(q_{_\bot})}{J(i)q_{_{_P}}+\alpha_iM+\omega_{_{iP}}-i\epsilon},
\end{eqnarray}
with
\begin{eqnarray}
\omega_{_{iP}}=\sqrt{m_i-q_{_\bot}^2},~~~
\Lambda^{\pm}_{iP}(q_{_\bot})=\frac{1}{2\omega_{_{iP}}}\left[\frac{\slashed{P}}{M}\omega_{_{iP}}\pm J(i)(m_i+\slashed{q}_{\bot})\right],
\end{eqnarray}
where $J(1)=-1$, $J(2)=1$, and the energy projection operator $\Lambda^{\pm}$ satisfies the relation:
\begin{eqnarray}\label{proje}
\Lambda^{+}_{iP}(q_{_\bot})+\Lambda^{-}_{iP}(q_{_\bot})=\frac{\slashed{P}}{M},~
\Lambda^{\pm}_{iP}(q_{_\bot})\frac{\slashed{P}}{M}\Lambda^{\pm}_{iP}(q_{_\bot})=\Lambda^{\pm}_{iP}(q_{_\bot}),~
\Lambda^{\pm}_{iP}(q_{_\bot})\frac{\slashed{P}}{M}\Lambda^{\mp}_{iP}(q_{_\bot})=0.
\end{eqnarray}

Using the energy projection operators, we define,
\begin{eqnarray}
&&\varphi_P^{\pm\pm}(q_{_\bot})=\Lambda^\pm_{1P}(q_{_\bot})\frac{\slashed{P}}{M}\varphi_{_P}(q_{_\bot})\frac{\slashed{P}}{M}\Lambda^\pm_{2P}(q_{_\bot}),
\end{eqnarray}
then the wave function can be decomposed as:
\begin{eqnarray}
&&\varphi_{_P}(q_{_\bot})=\varphi_P^{++}(q_{_\bot})+\varphi_P^{+-}(q_{_\bot})+\varphi_P^{-+}(q_{_\bot})+\varphi_P^{--}(q_{_\bot}).
\end{eqnarray}

Substitute the expression of the propagator Eq.(\ref{propa}) into the BS Eq.(\ref{bse}), integrate both sides with respect to $q_{_P}$, and apply the residue theorem. We then obtain the Salpeter equation \cite{Salpeter}:
\begin{eqnarray}
&&\varphi_{_P}(q_{_\bot})=\frac{\Lambda_{2P}^+(q_{_\bot})\eta(q_{_\bot})\Lambda_{1P}^+(q_{_\bot})}{M-\omega_{1P}-\omega_{2P}}-\frac{\Lambda_{2P}^-(q_{_\bot})\eta(q_{_\bot})\Lambda_{1P}^-(q_{_\bot})}{M+\omega_{1P}+\omega_{2P}}.
\end{eqnarray}
Applying the relations of the projection operators in Eq.(\ref{proje}), we obtain another  equivalent representation of the Salpeter equation \cite{Salpeter}:
\begin{eqnarray}
&&(M-\omega_{1P}-\omega_{2P})\varphi_P^{++}(q_{_\bot})=\Lambda^+_{2P}(q_{_\bot})\eta(q_{_\bot})\Lambda^+_{1P}(q_{_\bot}),\nonumber\\
&&(M+\omega_{1P}+\omega_{2P})\varphi_P^{--}(q_{_\bot})=-\Lambda^-_{2P}(q_{_\bot})\eta(q_{_\bot})\Lambda^-_{1P}(q_{_\bot}),\nonumber\\
&&\varphi_P^{+-}(q_{_\bot})=\varphi_P^{-+}(q_{_\bot})=0.
\end{eqnarray}
The normalization condition for the Salpeter wave function is
\begin{eqnarray}
\int\frac{d^3q_{_\bot}}{(2\pi)^3}\left[\bar{\varphi}_P^{++}(q_{_\bot})\frac{\slashed{P}}{M}\varphi_P^{++}(q_{_\bot})\frac{\slashed{P}}{M}-\bar{\varphi}_P^{--}(q_{_\bot})\frac{\slashed{P}}{M}\varphi_P^{--}(q_{_\bot})\frac{\slashed{P}}{M}\right]=2M.
\end{eqnarray}

The Salpeter equation itself does not directly provide a representation of the wave function. We adopt a general relativistic wave function representation, which contains four (or eight) independent unknown radial wave functions, see Sec.IV. In our calculation, we employed the latter representation of the Salpeter equation, meaning we solved the Salpeter equation comprising four coupled sub-equations. With four equations and four unknowns, the system is exactly solvable. Furthermore, since the wave function contains gamma matrices, additional equations can be obtained by taking traces. This means that the case with eight independent unknown radial wave functions can also be solved normally. It is evident that, apart from employing the instantaneous approximation, we have not introduced any further approximations.}

\section{Semileptonic and Nonleptonic decay Formulas}
\begin{figure}[!htb]
\begin{minipage}[c]{1\textwidth}
\includegraphics[width=3in]{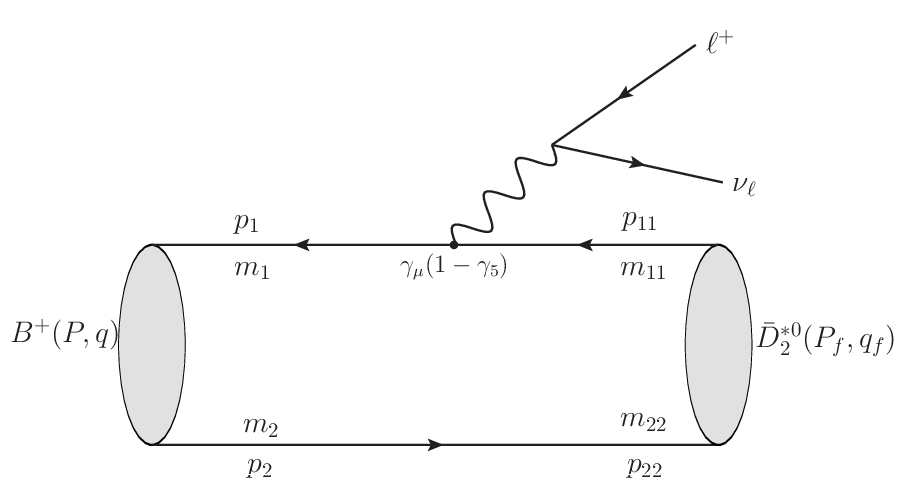}
\end{minipage}
\caption{Feynman diagram for  $B^{+}\rightarrow \bar{D}_2^{*0}\ell^+\nu_{\ell}. $ }
\label{BQfeynman}
\end{figure}

{The semileptonic decay $B^{+}\rightarrow D_2^{*0}\ell^+\nu_{\ell}$ is induced by the weak transition of $\bar{b}\to \bar{c}$, and the corresponding Feynman diagram is shown in Fig. \ref{BQfeynman}, with relevant physical quantities labeled, such as the external momenta $P$ and $P_f$ of the initial and final mesons, their internal momenta $q$ and $q_f$, etc.}

The transition amplitude for the decay $B^{+}\rightarrow D_2^{*0}\ell^+\nu_{\ell}$ can be written as
\begin{eqnarray}
T=\frac{G_F}{\sqrt{2}}V_{cb}\bar{u}_{\nu_{\ell}}\gamma^{\mu}(1-\gamma_{5})
v_{\ell}\langle{\bar{D}_2^{*0}}|J_{\mu}|{B^+}\rangle,
\end{eqnarray}
where $G_F$ is the Fermi weak coupling constant, $V_{cb}$ is the CKM matrix element,  $u_{\nu_{\ell}}$ is the spinor of the neutrino $\nu_{\ell}$, $v_{\ell}$ is the spinor of the antilepton $\ell^+$, and $J_{\mu}$ is the charged weak current responsible for the $\bar b \to \bar c$ transition, etc.

Unlike the leptonic part $\bar{u}_{\nu_{\ell}}\gamma^{\mu}(1-\gamma_{5})
v_{\ell}$, the calculation of hadronic transition matrix element $\left \langle  D_{2}^{*0} (P_f) |J_\mu |B^+(P)  \right \rangle$ is model-dependent. We will utilize the solutions of the instantaneous approximated BS equation, the relativistic wave functions, to provide a relativistic treatment.
In our method, the hadronic transition matrix element is expressed as an overlap integral over the wave functions of the initial and final mesons \cite{wgl3}. Since we solve the Salpeter equation rather than the BS equation, the obtained wave functions are consequently instantaneous. To apply this wave function, we further impose the instantaneous approximation on the transition amplitude, in this case, the hadronic matrix element can be written as \cite{wgl3}:
\begin{eqnarray}\label{transi}
 && \left \langle  D_{2}^{*0} (P_f) |J_\mu |B^+(P)  \right \rangle =\int\frac{d\vec{q}}{(2 \pi )^3} \mathrm{Tr}\left[\frac{\slashed P}{M} \varphi^{++}_P(\vec{q}) \gamma _\mu (1-\gamma _5)  \bar{\varphi}^{++}_{P_f} (\vec{q_f}) \right]\nonumber \\&&=K_1~\epsilon^*_{\mu P}+K_2~\epsilon^*_{P P} P_\mu +K_3~\epsilon^*_{P P} P_{f \mu}  +i~K_4~ \epsilon^{*\rho P} \varepsilon_{\rho P P_f \mu},
\end{eqnarray}
where $M$ is the mass of $B$ meson, $\epsilon_{\mu\nu}$ is the polarization tensor of $D_{2}^{*}$, $\varepsilon_{\mu\nu\alpha\beta}$ is the Levi-Civita symbol, $K_i$ ($i=1,2,3,4$) is the form factor,  $\varphi^{++}_P(\vec{q})$ and ${\varphi}^{++}_{P_f} (\vec{q_f})$ are the positive energy wave functions of the initial $B$ and final  $D_{2}^{*0} (P_f)$ with $\bar{\varphi}^{++}_{P_f}=\gamma_0 ({\varphi^{++}_{P_f}})^{\dagger} \gamma_0$, respectively. In Fig. \ref{BQfeynman}, quark 2 is a spectator, so we have $m_{22}=m_2$ and $p_{22}=p_{2}$ then the relation $\vec{q_f}=\vec{q}-\frac{m_2}{m_{11}+m_2}\vec{P_f}$ is obtained. In Eq. (\ref{transi}), we have ignored the negative energy wave functions and other similar terms which have very small contributions, and we have used some abbreviations, for example, $\epsilon^{*\rho\sigma} {P_{\sigma}}~ \varepsilon_{\rho\alpha\beta  \mu}P^{\alpha} {P_f}^{\beta}=\epsilon^{*\rho P} \varepsilon_{\rho P P_f \mu}$.

After taking the modulus squared of the transition amplitude, averaging over the spins of the initial meson and summing over polarizations of final state, we obtain:
\begin{eqnarray}
\sum{|T|^2=\frac{G^2_F}{2}V^2_{cb}\ell^{\mu\nu}h_{\mu\nu}},
\end{eqnarray}
where $\ell^{\mu\nu}$ represents the leptonic tensor, which can be expressed as $\ell^{\mu \nu} \equiv {\sum}\bar{\mu}_{\nu_\ell}\gamma^\mu(1-\gamma_5)\upsilon _\ell\bar{\upsilon }_\ell(1+\gamma_5)\gamma^\nu\mu_{\nu_\ell}$, and $h_{\mu\nu}$ denotes the hadronic tensor, it can be expressed as
\begin{eqnarray}
h_{\mu \nu} =&& {\sum}\left \langle B^+ |J_\nu^+ | \bar{D}_{2}^{*0} \right \rangle \left \langle \bar{D}_{2}^{*0}|J_\mu |B^+ \right \rangle
\nonumber \\=&&-\alpha g_{\mu \nu}+\beta_{++}\left(P+P_f\right)_{\mu}\left(P+P_f\right)_{\nu}+\beta_{+-}\left(P+P_f\right)_{\mu}\left(P-P_f\right)_{\nu} \nonumber \\
&&+\beta_{-+}\left(P-P_f\right)_{\mu}\left(P+P_f\right)_{\nu}+\beta_{--}\left(P-P_f\right)_{\mu}\left(P-P_f\right)_{\nu} \nonumber \\
&&+ i \gamma \varepsilon_{\mu \nu \rho \sigma}\left(P+P_f\right)^{\rho}\left(P-P_f\right)^{\sigma},
\end{eqnarray}
where the coefficients $\alpha$, $\beta_{\pm\pm}$ and $\gamma$ are functions of the form factors $K_i~ (i=1,2,3,4)$.

Thus, the differential decay rate of this exclusive process can be written as
\begin{eqnarray}
\frac{d^{2} \Gamma}{d x d y}= && \left|V_{cb}\right|^{2} \frac{G_{F}^{2} M^{5}}{32 \pi^{3}}\left\{\alpha \frac{\left(y-\frac{m_{\ell}^{2}}{M^{2}}\right)}{M^{2}}+2 \beta_{++} \right. \nonumber \\
&& \times{\left[2 x\left(1-\frac{M_f^{ 2}}{M^{2}}+y\right)-4 x^{2}-y+\frac{m_{\ell}^{2}}{4 M^{2}}\left(8 x+\frac{4 M_f^{2}-m_{\ell}^{2}}{M^{2}}-3 y\right)\right]} \nonumber \\
&& +\left(\beta_{+-}+\beta_{-+}\right) \frac{m_{\ell}^{2}}{M^{2}}\left(2-4 x+y-\frac{2 M_f^{2}-m_{\ell}^{2}}{M^{2}}\right)+ \beta_{--} \frac{m_{\ell}^{2}}{M^{2}}\left(y-\frac{m_{\ell}^{2}}{M^{2}}\right) \nonumber \\
&& \left.-\gamma\left[y\left(1-\frac{M_f^{2}}{M^{2}}-4 x+y\right)+\frac{m_{\ell}^{2}}{M^{2}}\left(1-\frac{M_f^{2}}{M^{2}}+y\right)\right]\right\},
\end{eqnarray}
where $x\equiv E_\ell/M,~y\equiv (P-P_f)^2/M^2$, $M_f$ is the masse of $\bar D_{2}^{*0}$, and {$m_\ell$ and $E_\ell$} are the mass and energy of the final charged lepton $\ell^+$, respectively.

\begin{figure}[!htb]
\begin{minipage}[c]{1\textwidth}
\includegraphics[width=3in]{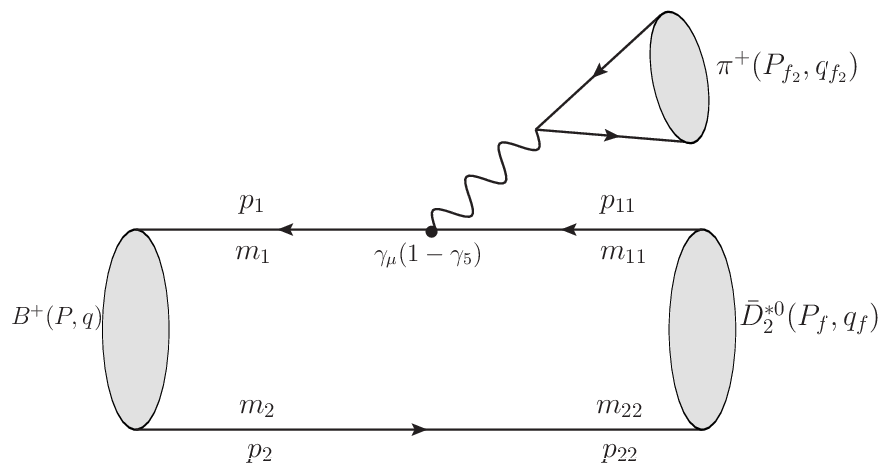}
\end{minipage}
\caption{Feynman diagram for  $B^{+}\rightarrow D_2^{*0}\pi^{+}. $ }
\label{Nfeynman}
\end{figure}
We also present the calculation for the nonleptonic decay
$B^+$$\rightarrow$$\bar{D}_2^{*0}\pi^+$, whose Feynman diagram is depicted in Fig. \ref{Nfeynman}. Adopting the naive factorization approach, the transition amplitude for the nonleptonic decay can be expressed as the product of the hadronic transition matrix element and decay constant:
\begin{eqnarray}
T=\frac{G_F}{\sqrt{2}}V_{cb}V^*_{ud}a_1\langle{\bar{D}_2^{*0}}|J_{\mu}|{B^+}
\rangle\langle{\pi^+}|J'^{\mu}|{0}\rangle,
\end{eqnarray}
where the $V_{ud}^*$ is also the CKM matrix element, $J'^{\mu}$ is the charged weak  current responsible for the $d \to u$ transition, $a_1=c_1+\frac{1}{N_c}c_2$, with $N_c$ indicating the number of color, and $c_1$ and $c_2$ being the energy-dependent Wilson coefficients. The matrix element $\langle{\pi^+}|J'^{\mu}|{0}\rangle$ is related to the decay constant, i.e., $\langle{\pi^+}|J'^{\mu}|{0}\rangle=if_{\pi}P^{\mu}_{\pi}$, where $f_{\pi}$ and $P^{\mu}_{\pi}$ ($=P^{\mu}_{f_2}$) are the decay constant and momentum of the $\pi$ meson.

\section{Relativistic wave functions}
We will solve the full Salpeter equation. However, as the equation itself does not constrain the representation of the wave function, the specific form of the wave function must be externally input into the Salpeter equation, with its radial wave function being the quantity to be solved. To distinguish particles, we have adopted the naming convention based on the principal quantum number $n$ and orbital angular momentum $L$, i.e., labeling particle as $n^{2S+1}L_J$ (or simply $nL$). Nevertheless, this classification scheme, such as $S$-wave ($L=0$), $P$-wave ($L=1$), $D$-wave ($L=2$), and $F$-wave ($L=3$), originating from atomic physics, is inherently non-relativistic. If the representation remains non-relativistic, the wave function obtained by solving the BS equation would also retains a non-relativistic character. Therefore, we will construct the wave function expression based on the
$J^P$ quantum number of the particle ( not the $n^{2S+1}L_J$) to ensure the relativistic nature, where $P=(-1)^{L+1}$ is the parity and $J$ the total angular momentum of the meson. Due to this method, the wave function we propose is no longer pure wave, but an admixture of multiple wave components. Furthermore, the resulting $D^*_2(nF)$ wave function is not merely a $P-F$ mixture, but rather a $P-D-F$ mixture.

\subsection{Wave function of the $0^-$ state}
{In general, the wave function of a pseudoscalar can be expressed as a sum of 8 terms. However, under the instantaneous approximation ${P}\cdot{q_{_{\bot}}}=0$, four terms are eliminated.} The general relativistic wave function for a $0^-$ state can be expressed as \cite{0-}:
\begin{eqnarray}\label{0-}
\varphi^{0^{-}}_{P}(q_{_{\bot}})=&&\left(h_{1}\slashed{P}+h_{2}M+h_{3}\slashed{q}_{{\bot}}+h_{4}\frac{\slashed{P} \slashed{q}_{{\bot}} }{M}\right)\gamma_5,
\end{eqnarray}
where the unknown radial wave function $h_i(i=1,2,3,4)$ is a function of $-q^2_{_{\bot}}$ ($=\vec{q}^2$ in the CMS of $P$), and its numerical value is obtained by solving the full Salpeter equation. The Salpeter equation shows us that not all $h_is$ are independent, they have the following relations \cite{0-}:
$$
h_3=\frac{h_2M(\omega_{_{2}}-\omega_{_{1}})}{m_1\omega_{_{2}}+m_2\omega_{_{1}}},\quad\quad\quad\quad
h_4=-\frac{h_1M(\omega_{_{2}}+\omega_{_{1}})}{m_1\omega_{_{2}}+m_2\omega_{_{1}}},
$$ where $\omega_{_{1}}=\sqrt{m^2_1-q^2_{_{\bot}}}$ and $\omega_{_{2}}=\sqrt{m^2_2-q^2_{_{\bot}}}$ are the energies of quarks, respectively.

It is straightforward to verify that each term of the wave function in Eq. (\ref{0-}) possesses $0^-$ quantum number. Moreover, the terms involving $h_1$ and $h_2$ correspond to $S$-waves, representing non-relativistic contributions, while those containing $h_3$ and $h_4$ correspond to $P$-waves, accounting for relativistic correction contributions. The positive-energy wave function for a $0^-$ meson is expressed as a functional of the wave function in Eq. (\ref{0-}), whose specific form is provided in the Appendix A. The details of solving the full Salpeter equation for pseudoscalar meson to obtain the radial wave functions are not provided here, interested readers are referred to Ref. \cite{0-}, we only show the adopted interaction potential in Appendix B.

\subsection{Wave function of the $2^+$ state}
In the condition of instantaneous approximation, the general relativistic wave function for a \(2^{+}\) state can be expressed as \cite{wgl5}:
\begin{eqnarray}\label{2+eq}
\varphi^{2^{+}}_P(q_{_\bot})=&&\varepsilon_{\mu\nu}q_{\bot}^{\mu}q_{\bot}^{\nu}\ \left(f_{1}+\frac{\slashed{P}}{M}f_{2}+\frac{\slashed{q}_{\bot}}{M}f_{3}+
\frac{\slashed{q}_{\bot}\slashed{P}}{M^{2}}f_{4}\right)+\nonumber\\
&& + \varepsilon_{\mu\nu}q_{\bot}^{\nu}\gamma^{\mu}\left( Mf_{5}+\slashed{P}f_{6}+\slashed{q}_{\perp}f_{7}+\frac{i}{M}f_{8}\epsilon_{\mu\alpha\beta\gamma}P^{\alpha} q_{\perp}^{\beta}\gamma^{\gamma}\gamma_{5}\right),
\end{eqnarray}
where $\varepsilon_{\mu\nu}$=$\varepsilon_{\nu\mu}$ is the polarization tensor of the $2^+$ meson;  $f_{i}(i=1,2,......,8)$, being a function of $-{q}_{\bot}^2$, is the radial wave function of the meson. Similarly, only four $f_{i}s$ are independent, and there exists the following relationships between them \cite{wgl5},
$$
f_{1}=\frac{2Mf_{5}\omega_{2}}{m_{1}\omega_{2}+m_{2}\omega_{1}}+\frac{q^{2}_{_\bot}f_{3}(\omega_{1}+\omega_{2})}{M(m_{1}\omega_{2}+m_{2}\omega_{1})},~~~
f_{7}=\frac{f_{5}M(\omega_{1}-\omega_{2})}{(m_{1}\omega_{2}+m_{2}\omega_{1})},$$
$$f_{2}=\frac{2Mf_{6}\omega_{2}}{m_{1}\omega_{2}+m_{2}\omega_{1}}+\frac{q^{2}_{_\bot}f_{4}(\omega_{1}-\omega_{2})}{M(m_{1}\omega_{2}+m_{2}\omega_{1})},~~~
f_{8}=\frac{f_{6}M(\omega_{1}+\omega_{2})}{(m_{1}\omega_{2}+m_{2}\omega_{1})}.$$

It can be verified that the terms containing $f_5$ and $f_6$ which involve a single $q_{\bot}$ correspond to $P$-wave, and those with $f_1$, $f_2$, $f_7$ and $f_8$ including two $q_{\bot}$s are $D$-wave, while the $f_3$ and $f_4$ terms involve three $q_{\bot}$s, contain both $F$-wave and $P$-wave, representing a $P-F$ mixture. For details, the $f_3$ and $f_4$ terms contribute to $P$-wave as
\begin{eqnarray}
\frac{2}{5}
\epsilon _{\mu\nu}q_{\bot }^{\mu }\gamma^vq_{\bot }^{2 }(\frac{f_3}{M}-\frac{\slashed P}{M^2} f_4 ),
\end{eqnarray}
and contribute to $F$-wave as
\begin{eqnarray}
\epsilon _{\mu\nu}q_{\bot }^{\mu }q_{\bot }^{\nu}
(\frac{\slashed {q}_\bot}{M}f_3+\frac{\slashed P \slashed {q}_\bot  }{M^2} f_4)-\frac{2}{5}
\epsilon _{\mu\nu}q_{\bot }^{\mu }\gamma^{\nu} q_{\bot }^{2 }(\frac{f_3}{M}-\frac{\slashed P}{M^2} f_4 ).
\end{eqnarray}

Eq. (\ref{2+eq}) provides the general wave function representation for $2^+$ meson, where the radial wave functions are determined by solving the dynamic Salpeter equation. As a relativistic two-body bound-state equation, its solutions yield not only the eigenstates of $1P$, $2P$ and $3P$ $2^+$ states, but also those of $1F$ and $2F$ states sharing the same quantum number of $2^+$. The details will be discussed in the next section.
And the expression of the positive energy wave function of a $2^+$ meson is also shown in the Appendix A.

In comparison with non-relativistic wave function representations \cite{top}, the relativistic wave functions for the $0^-$ and $2^+$ states presented in this work demonstrate that the dominant partial waves, such as the $S$-wave in the $0^-$ state and the $P$-wave in the $1P$ and $2P$ $2^+$ states, provide the non-relativistic contributions. Meanwhile, the subordinate partial waves, such as the $P$-wave in the $0^-$ state and the $D$- and $F$-waves in the $1P$ and $2P$ $2^+$ states, contribute the relativistic corrections. Thus, the mixing phenomenon arises naturally within relativistic theory, and the magnitude of relativistic corrections determines the strength of mixing.

\section{Results and Discussions}\label{5}
In our calculations, we adopt the following parameters: $m_b$=4.96 \rm{GeV}, $m_c$=1.68 \rm{GeV}, $m_d$=0.356 \rm{GeV}, $m_u$=0.35 \rm{GeV}, and $a_1$=1.02. Additionally, the masses of mesons and leptons, the CKM matrix elements, and other relevant values are same as those listed in the Particle Data Group (PDG) tables \cite{1P5}.

\subsection{Solutions of Salpeter equation and numerical values of $2^+$ wave functions}
\begin{figure}[!htb]
\begin{minipage}[c]{1\textwidth}
\includegraphics[width=2.1in]{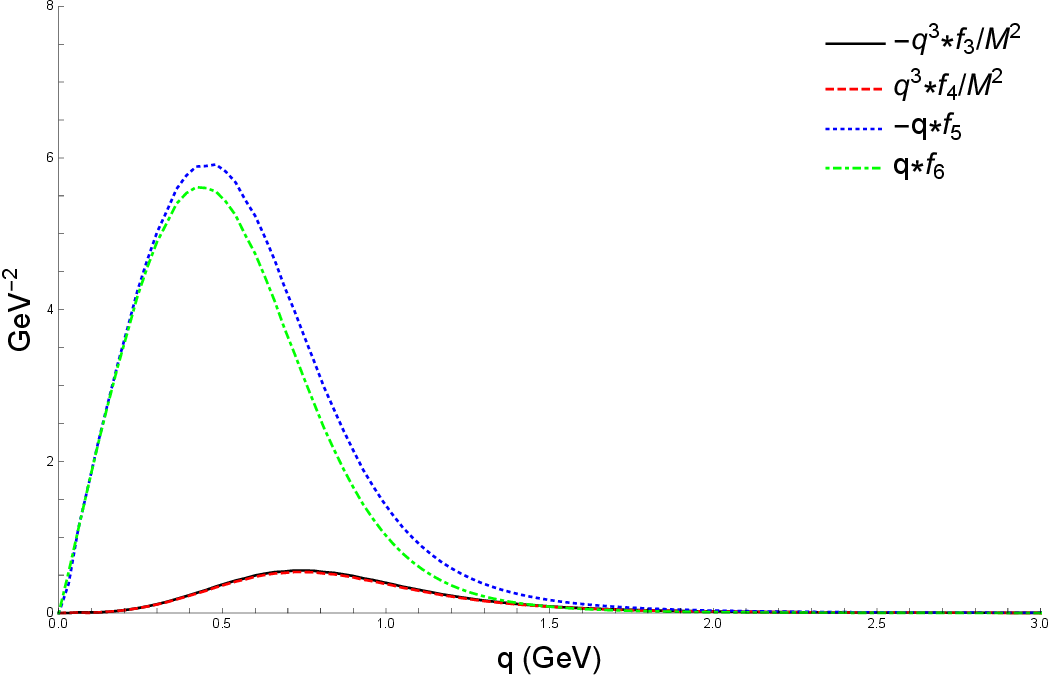}
\includegraphics[width=2.1in]{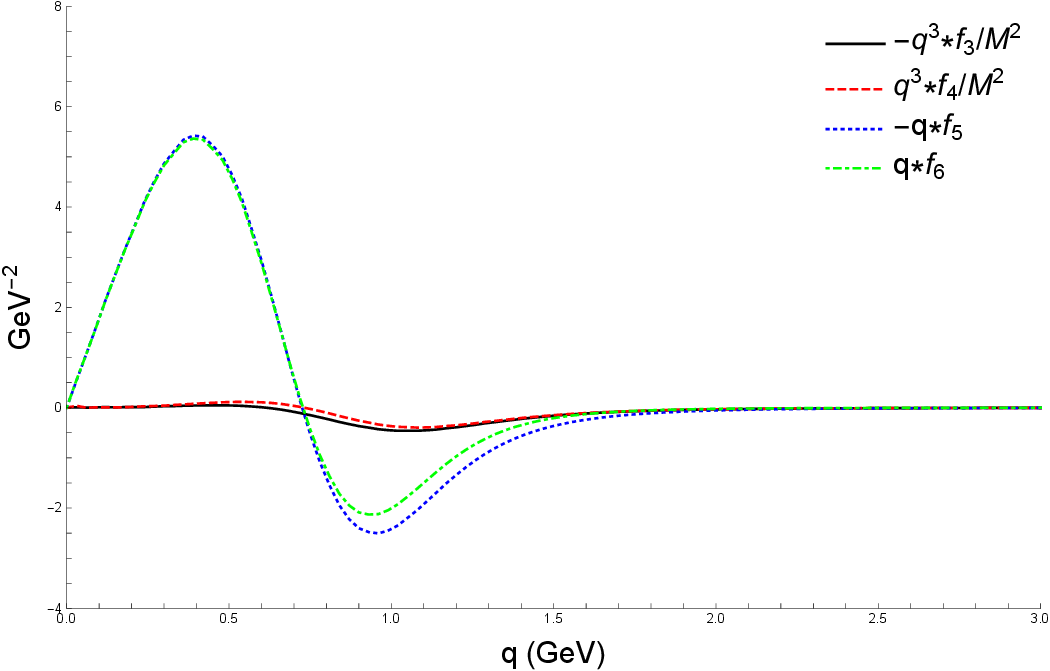}
\includegraphics[width=2.1in]{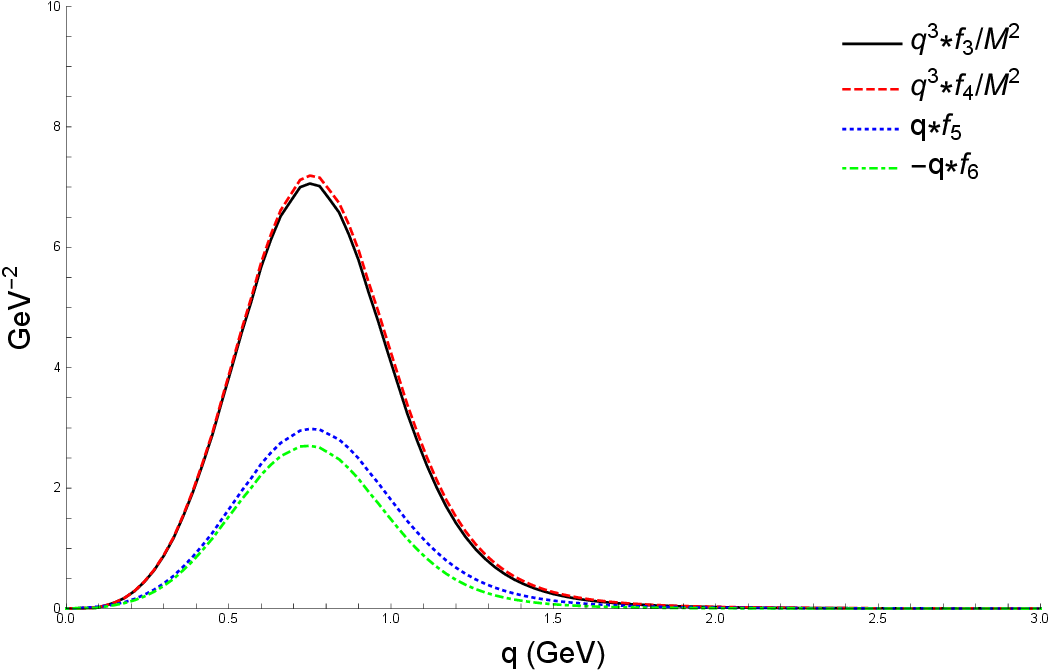}
\end{minipage}
\caption{The radial wave functions of the $2^+$ mesons $\bar{D}_2^{*0}(2460)$ (left), $\bar{D}_2^{*0}(2P)$ (middle) and $\bar{D}_2^{*0}(1F)$ (right), where $q\equiv |\vec{q}|$.}
\label{2+}
\end{figure}

With the representation of the wave function given in Eq. (\ref{2+eq}), we substitute it into the complete Salpeter equation for solving \cite{wgl5}. As an eigenvalue equation, we simultaneously obtain both the numerical solutions for the radial wave functions and the mass spectrum of mesons. The obtained solutions correspond orderly to the $1P$, $2P$, $1F$, $3P$, and $2F$ states, and so on. The lowest-mass of the first solution, 2.46 GeV, is our input. The independent radial wave functions for the first three solutions are plotted in Fig. \ref{2+}. As can be seen, the $P$-wave components $f_5$ and $f_6$ dominate in the first solution (left), and the wave function exhibits no nodes. Therefore, this solution corresponds to the ground $2^+$ state, i.e., the $1P$ meson $\bar{D}_2^{*0}(2460)$.
The second solution (middle) with mass 2.99 GeV exhibits a wave function containing one node while maintaining dominant $P$-wave characteristics, corresponding to the first excited state ($2P$ meson $\bar{D}_2^{*0}(2P)$). The third solution (right) with mass 3.09 GeV, showing a nodeless wave function but dominated by $F$-wave components ($f_3$ and $f_4$), corresponds to the $1F$ meson $\bar{D}_2^{*0}(1F)$ (second excited state). The fourth and fifth solutions correspond to the 3P and 2F states, whose wave functions are plotted in Fig. 4 without elaborating further on the details.

\begin{figure}[!htb]
\begin{minipage}[c]{1\textwidth}
\includegraphics[width=2.1in]{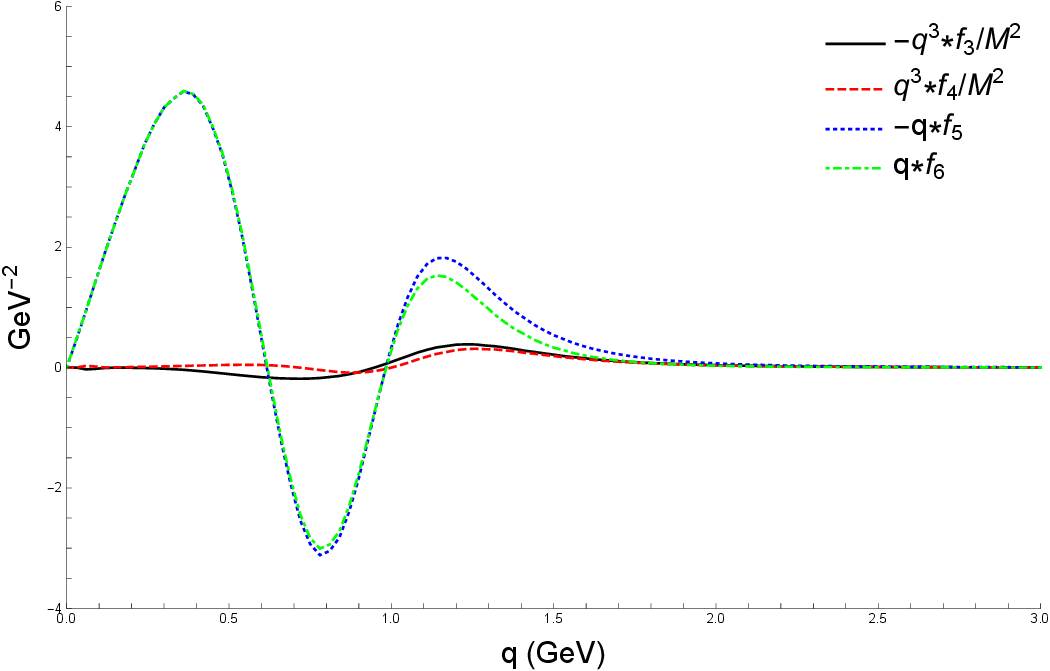}
\includegraphics[width=2.1in]{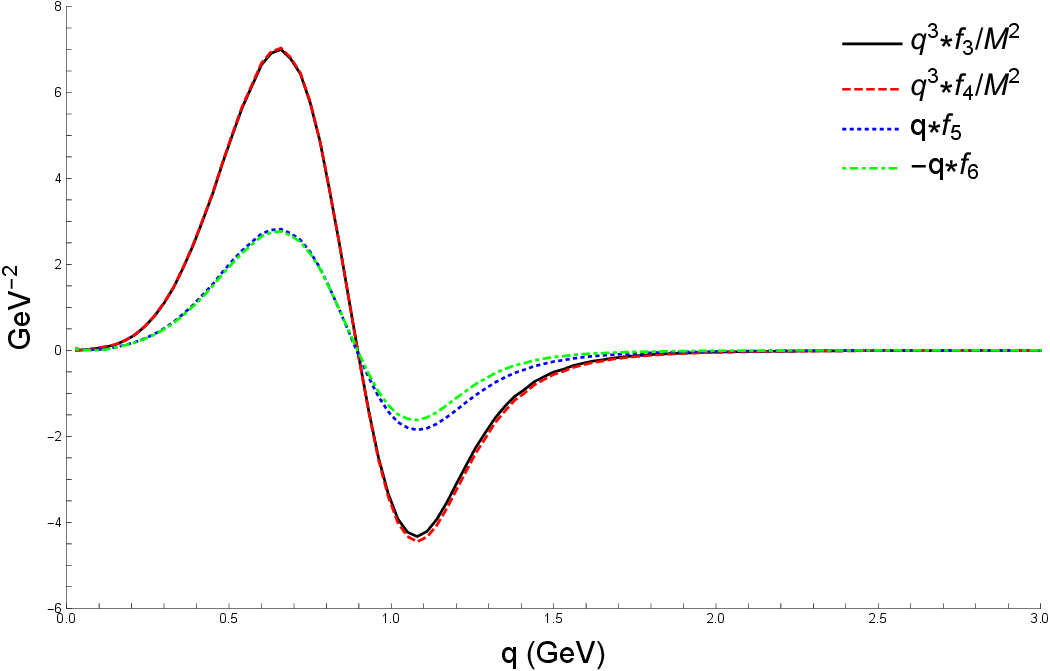}
\end{minipage}
\caption{The radial wave functions of the $2^+$ mesons $\bar{D}_2^{*0}(3P)$ (left), and $\bar{D}_2^{*0}(2F)$ (right), where $q\equiv |\vec{q}|$.}
\label{2+2}
\end{figure}

As introduced earlier regarding wave function representation, we pointed out that the relativistic wave function for the $2^+$ state contains not only $P$- and $F$-wave components but also a $D$-wave component. Having obtained the numerical solutions of the radial wave functions, we compute the ratios between these partial waves, $P:D:F$, with results presented in Table \ref{2++ partial wave} for $nP$ ($n=1,2,3$) and $mF$ ($m=1,2$) mesons. As shown in the table, the wave functions of the $nP$ states are dominated by the $P$-wave, with a significant $D$-wave component and a minor $F$-wave admixture. The $P$-wave provides the non-relativistic contribution, while the $D$- and $F$-waves give relativistic corrections. In contrast, the wave function of the $mF$ mesons are primarily composed of the $F$-wave (providing the non-relativistic contribution), with a notable $D$-wave component and a small $P$-wave admixture. Among these, the $D$-wave furnishes the dominant relativistic correction.
\begin{table}[ht]
\centering \caption{Ratios between the partial waves in the wave functions of  $2^{+}$ mesons} \label{2++ partial wave}
\setlength{\tabcolsep}{6pt}
\renewcommand{\arraystretch}{1}
\begin{tabular}{|c|c|c|c|c|c|c|}
\hline
        $D_2^{*0}$     &   $1P$           &    $2P$             &    $1F$          &    $3P$          & $2F$\\ \hline
        $P:D:F$        & 1:~0.367:~0.063  &  1:~0.458:~0.078    &  0.074:~0.550:~1  &  1:~0.538:~0.071 &  0.066:~0.596:~1   \\ \hline
\end{tabular}
\end{table}

\subsection{Validity of the Method}
Before presenting the results for $2P$, $1F$ and $3P$ mesons, etc, we first validate the effectiveness of our method. Currently, the experimental data are available for both semileptonic and nonleptonic decays of $B$ mesons to the $1P$ state $D^*_2(2460)$. For the semileptonic decays, the experimental results are \cite{1P5}:
\begin{eqnarray}
&&\mathcal{B}\left(B^{+} \rightarrow \bar{D}_{2}^{*}(2460)^0 \ell^{+} \nu_{\ell} \right) \mathcal{B}\left(\bar{D}_{2}^{*}(2460)^0 \to D^{-} \pi^{+}\right) =  (1.53 \pm 0.16) \times 10^{-3}, \nonumber\\
&&\mathcal{B}\left(B^{0} \to {D}_{2}^{*}(2460)^- \ell^{+} \nu_{\ell} \right) \mathcal{B}\left(D_{2}^{*}(2460)^- \to  \bar{D}^{0} \pi^{-}\right)  = (1.21 \pm 0.33) \times 10^{-3}.\label{fenzhibi}
\end{eqnarray}
The branching ratios for the semileptonic $B$ decays obtained via our method are
\begin{eqnarray}
&&\mathcal{B}\left(B^{+} \rightarrow \bar{D}_{2}^{*}(2460)^0 \ell^{+} \nu_{\ell} \right)  = (2.67^{+1.05}_{-0.80})\times10^{-3}, \nonumber\\
&&\mathcal{B}\left(B^{0} \rightarrow {D}_{2}^{*}(2460)^- \ell^{+} \nu_{\ell} \right)      = (2.59^{+1.03}_{-0.78})\times10^{-3},\label{fenzhibi2}
\end{eqnarray}where the theoretical uncertainties are obtained by varying all the input parameters simultaneously within $\pm5\%$ of their central values.
In our previous paper \cite{1P}, we calculated the OZI-allowed strong decays of ${D}_{2}^{*}(2460)$ and obtained the branching ratios $\mathcal{B}(\bar{D}_{2}^{*}(2460)^0 \to D^{-} \pi^{+})=\mathcal{B}(D_{2}^{*}(2460)^- \to  \bar{D}^{0} \pi^{-})=44.5\char\%$. Using these values, we obtain:
\begin{eqnarray}
&&\mathcal{B}\left(B^{+} \rightarrow \bar{D}_{2}^{*}(2460)^0 \ell^{+} \nu_{\ell} \right) \mathcal{B}\left(\bar{D}_{2}^{*}(2460)^0 \rightarrow D^{-} \pi^{+}\right) =  (1.19^{+0.47}_{-0.36}) \times 10^{-3}, \nonumber\\
&&\mathcal{B}\left(B^{0} \rightarrow {D}_{2}^{*}(2460)^- \ell^{+} \nu_{\ell} \right) \mathcal{B}\left(D_{2}^{*}(2460)^- \rightarrow  \bar{D}^{0} \pi^{-}\right)    =  (1.15^{+0.46}_{-0.34}) \times 10^{-3}.\label{fenzhibi3}
\end{eqnarray}
It can be seen that the central value of our result for the $B^+$ meson is smaller than the experimental value, but it is still consistent within the error margin. The result for the $B^0$ meson, however, shows excellent agreement with the experimental data.

Here we must point out that the error in Eq. (\ref{fenzhibi3}) originates solely from the semileptonic decay $B \rightarrow {D}_{2}^{*}(2460)\ell \nu_{\ell}$, while the uncertainty from the strong decay ${D}_{2}^{*}(2460) \rightarrow D \pi$ is not included. This omission is for two reasons: first, the uncertainty of ${D}_{2}^{*}(2460) \rightarrow D \pi$ was not calculated in Ref. \cite{1P}; second, the purely theoretical branching ratio $\mathcal{B}\left({D}_{2}^{*}(2460) \rightarrow D \pi\right)$ significantly reduces the uncertainty, meaning the theoretical error of $\mathcal{B}\left({D}_{2}^{*}(2460) \rightarrow D \pi\right)$ is relatively small. For the same reason, the uncertainties of the branching ratios involving strong decays of all ${D}_{2}^{*}$ states discussed later are not considered.

For the nonleptonic decay processes, the experimental data are \cite{1P5}:
\begin{eqnarray}
&&\mathcal{B}\left(B^{+} \rightarrow \bar{D}_{2}^{*}(2460)^0 \pi^{+} \right) \mathcal{B}\left(\bar{D}_{2}^{*}(2460)^0 \rightarrow D^{-} \pi^{+}\right) =  (3.56 \pm 0.24) \times 10^{-4}, \nonumber\\
&&\mathcal{B}\left(B^{0} \rightarrow D_{2}^{*}(2460)^- \pi^{+} \right) \mathcal{B}\left(D_{2}^{*}(2460)^- \rightarrow {D}^{0} \pi^{-}\right)  = (2.38 \pm 0.16) \times 10^{-4},\label{fenzhibi4}
\end{eqnarray}
while our theoretical results
 \begin{eqnarray}
&&\mathcal{B}\left(B^{+} \rightarrow \bar{D}_{2}^{* 0} \pi^{+} \right) \mathcal{B}\left(\bar{D}_{2}^{* 0} \rightarrow D^{-} \pi^{+}\right) =(2.99^{+1.01}_{-0.74}\pm0.30\pm0.30) \times 10^{-4}, \nonumber\\
&&\mathcal{B}\left(B^{0} \rightarrow \bar{D}_{2}^{*-} \pi^{+} \right) \mathcal{B}\left(D_{2}^{*-} \rightarrow {D}^{0} \pi^{-}\right)       =(2.94^{+0.97}_{-0.77}\pm0.29\pm0.29) \times 10^{-4},
\end{eqnarray} essentially agree with the experimental data. The consistency between our results for both semileptonic and nonleptonic decays of $B$ mesons demonstrates the validity of our method. Compared to semi-leptonic decays, the theoretical error in non-leptonic decays has two additional sources, originating from the Wilson coefficient $a_1$ and the decay constant of the pion, respectively. Therefore, in the above expression, the first error still comes from the parameters of the potential model, while the second and third errors are obtained by varying $a_1$ and $f_{\pi}$ within a range of $\pm5\%$ of their central values.

\subsection{Results of Semileptonic and Nonleptonic Decays}

Our results with errors for semileptonic decays are summarized in Table \ref{II}. It can be seen that the branching ratios of semileptonic decays $B\to{D}_2^{*}(nP)\ell^+\nu_{\ell}$ ($n=2,3$) and $B\to{D}_2^{*}(mF)\ell^+\nu_{\ell}$ ($m=1,2$) are significantly smaller than that of $B\to{D}_2^{*}(2460)\ell^+\nu_{\ell}$. We present their ratios to the branching ratio of the $1P$ final state, which serves both to compare their magnitudes and to largely cancel out the model dependence. The ratios are:
\begin{eqnarray}
&&\frac{\Gamma[B^{+}\rightarrow \bar{D}_2^{*0}(2P)\ell^+\nu_{\ell}]}{\Gamma[B^+ \rightarrow\bar D_2^{\* 0}(2460)\ell^+\nu_{\ell}]}=(2.62^{+0.32}_{-0.18})\times10^{-2}, \nonumber\\
&&\frac{\Gamma[B^{+}\rightarrow \bar{D}_2^{*0}(1F)\ell^+\nu_{\ell}]}{\Gamma[B^+ \rightarrow\bar D_2^{\* 0}(2460)\ell^+\nu_{\ell}]}=(8.70^{+1.90}_{-1.98})\times10^{-4},\nonumber\\
&&\frac{\Gamma[B^{+}\rightarrow \bar{D}_2^{*0}(3P)\ell^+\nu_{\ell}]}{\Gamma[B^+ \rightarrow\bar D_2^{\* 0}(2460)\ell^+\nu_{\ell}]}=(5.43^{+0.32}_{-0.15})\times10^{-4}, \nonumber\\
&&\frac{\Gamma[B^{+}\rightarrow \bar{D}_2^{*0}(2F)\ell^+\nu_{\ell}]}{\Gamma[B^+ \rightarrow\bar D_2^{\* 0}(2460)\ell^+\nu_{\ell}]}=(2.45^{+0.55}_{-0.52})\times10^{-4}
\end{eqnarray}
The suppression of the $2P$, $3P$, and $2F$  final states arises from the nodal structure in their wave function, where contributions from wave function segments before and after the node cancel each other out. The small branching ratios for the $1F$ and $2F$ final states are also attributed to the small overlap between the initial-state $S$-wave and final-state $F$-wave wave functions, i.e., their small overlap integral.

\begin{table}[!htb]
\caption{Decay widths and branching ratios of semileptonic decays of $B$ to $2^+$ charmed mesons.}
\begin{tabular}{cccccccc}
\hline
&decay channel&~~~~~&Decay width~(\rm{MeV})&~~~~~&${Br}$& ~~~~~   \\ \hline
&$B^{+}\rightarrow \bar{D}_2^{*0}(2460)\ell^+\nu_{\ell}$&&$(1.07^{+0.42}_{-0.32})\times10^{-12}$&&$(2.67^{+1.05}_{-0.80})\times10^{-3}$&  \nonumber\\
&$B^{0}\rightarrow D_2^{*-}(2640)\ell^+\nu_{\ell}      $&&$(1.09^{+0.43}_{-0.33})\times10^{-12}$&&$(2.59^{+1.03}_{-0.78})\times10^{-3}$&   \nonumber\\
&$B^{+}\rightarrow \bar{D}_2^{*0}(2P)\ell^+\nu_{\ell}  $&&$(2.81^{+1.78}_{-1.00})\times10^{-14}$&&$(7.00^{+4.40}_{-2.49})\times10^{-5}$&  \nonumber\\
&$B^{0}\rightarrow D_2^{*-}(2P)\ell^+\nu_{\ell}        $&&$(2.55^{+2.31}_{-0.63})\times10^{-14}$&&$(6.07^{+5.53}_{-1.50})\times10^{-5}$&   \nonumber\\
&$B^{+}\rightarrow \bar{D}_2^{*0}(1F)\ell^+\nu_{\ell}  $&&$(9.31^{+6.79}_{-4.08})\times10^{-16}$&&$(2.32^{+1.69}_{-1.02})\times10^{-6}$&  \nonumber\\
&$B^{0}\rightarrow D_2^{*-}(1F)\ell^+\nu_{\ell}        $&&$(9.48^{+6.82}_{-4.13})\times10^{-16}$&&$(2.25^{+1.16}_{-0.98})\times10^{-6}$&   \nonumber\\
&$B^{+}\rightarrow \bar{D}_2^{*0}(3P)\ell^+ \nu_{\ell} $&&$(5.81^{+2.82}_{-1.79})\times10^{-16}$&&$(1.44^{+0.71}_{-0.44})\times10^{-6}$&  \nonumber\\
&${B}^{0}\rightarrow D_2^{*-}(3P)\ell^+ \nu_{\ell}     $&&$(7.04^{+1.81}_{-3.13})\times10^{-16}$&&$(1.67^{+0.44}_{-0.74})\times10^{-6}$&  \nonumber\\
&$B^{+}\rightarrow \bar{D}_2^{*0}(2F)\ell^+ \nu_{\ell} $&&$(2.62^{+2.67}_{-2.45})\times10^{-16}$&&$(6.53^{+6.67}_{-6.11})\times10^{-7}$&  \nonumber\\
&${B}^{0}\rightarrow D_2^{*-}(2F)\ell^+ \nu_{\ell}     $&&$(2.78^{+2.74}_{-2.47})\times10^{-16}$&&$(6.62^{+6.48}_{-5.88})\times10^{-7}$&  \nonumber\\
\hline
&$B^+ \rightarrow\bar D_2^{*0}(2460)\tau^+ \nu_{\tau}$&&$(0.48^{+0.62}_{-0.30})\times10^{-13}$&&$(1.19^{+1.55}_{-0.75})\times10^{-4}$&  \nonumber\\
&$B^{0}\rightarrow D_2^{*-}(2460)\tau^+ \nu_{\tau}   $&&$(0.49^{+0.64}_{-0.31})\times10^{-13}$&&$(1.15^{+1.97}_{-0.72})\times10^{-4}$&  \nonumber\\
&$B^+ \rightarrow\bar D_2^{*0}(2P)\tau^+ \nu_{\tau}  $&&$(0.86^{+2.48}_{-0.85})\times10^{-16}$&&$(2.14^{+6.19}_{-1.76})\times10^{-7}$&  \nonumber\\
&$B^{0}\rightarrow D_2^{*-}(2P)\tau^+ \nu_{\tau}     $&&$(0.73^{+2.81}_{-0.71})\times10^{-16}$&&$(1.74^{+6.71}_{-1.35})\times10^{-7}$&  \nonumber\\
&$B^+ \rightarrow\bar D_2^{*0}(1F)\tau^+ \nu_{\tau}  $&&$(3.06^{+6.94}_{-2.37})\times10^{-18}$&&$(0.76^{+1.73}_{-0.59})\times10^{-8}$&  \nonumber\\
&$B^{0}\rightarrow D_2^{*-}(1F)\tau^+ \nu_{\tau}     $&&$(3.21^{+7.09}_{-2.49})\times10^{-18}$&&$(0.76^{+1.69}_{-0.59})\times10^{-8}$& \\
&$B^{+}\rightarrow \bar{D}_2^{*0}(3P)\tau^+\nu_{\tau}$&&$(1.11^{+5.93}_{-0.98})\times10^{-19}$&&$(0.28^{+1.48}_{-0.25})\times10^{-9}$&  \nonumber\\
&${B}^{0}\rightarrow D_2^{*-}(3P)\tau^+ \nu_{\tau}   $&&$(1.39^{+5.81}_{-1.33})\times10^{-19}$&&$(0.33^{+1.39}_{-0.32})\times10^{-9}$&  \nonumber\\
&$B^{+}\rightarrow \bar{D}_2^{*0}(2F)\tau^+\nu_{\tau}$&&$(0.39^{+2.50}_{-0.34})\times10^{-19}$&&$(0.98^{+6.23}_{-0.81})\times10^{-10}$&  \nonumber\\
&${B}^{0}\rightarrow D_2^{*-}(2F)\tau^+ \nu_{\tau}   $&&$(0.39^{+2.34}_{-0.36})\times10^{-19}$&&$(0.92^{+5.58}_{-0.85})\times10^{-10}$&  \\
\hline
\end{tabular}\\
\label{II}
\end{table}

From Table \ref{II}, we note that the relative error of the semileptonic decay  $B\to{D}_2^{*}(2460)\ell^+\nu_{\ell}$ is the smallest. The higher the excited state, the larger the relative error becomes. The main reason for this phenomenon is that the wave functions of excited states contain nodes, and shifts in the node positions due to parameter changes significantly impact the results. Additionally, due to the large mass of highly excited states, the relatively small phase space of this process makes it more sensitive to parameters. The same conclusion applies to processes with $\tau$ leptons in the final state, resulting in much larger relative errors in $\tau$ lepton channel compared to the corresponding electron channel.

\begin{figure}[!htb]
\begin{minipage}[c]{1\textwidth}
\includegraphics[width=2.1in]{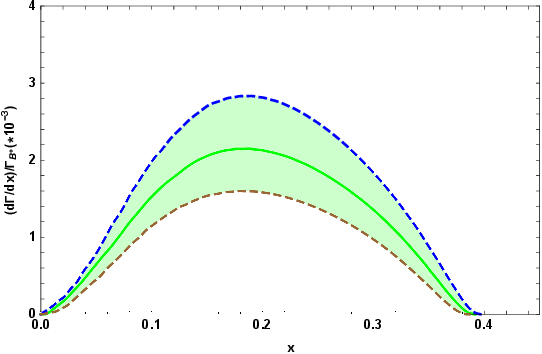}
\includegraphics[width=2.1in]{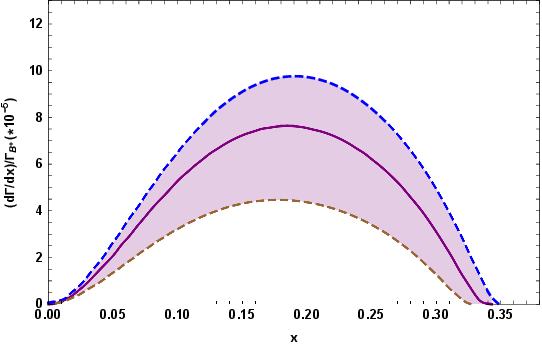}
\includegraphics[width=2.1in]{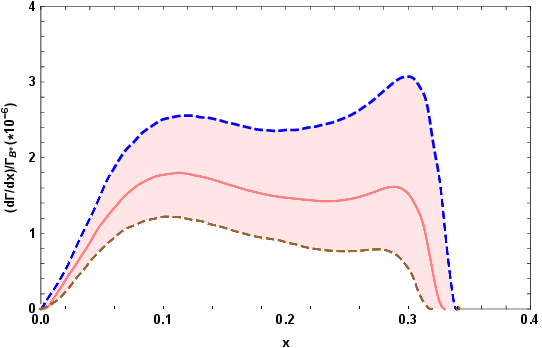}
\end{minipage}
\caption{Electron energy spectra of $(1/\Gamma)(d\Gamma/dx)$ for $B^{+}\to \bar{D}_2^{*0}(2460)\ell^+\nu_{\ell}$ (left), $B^{+}\to \bar{D}_2^{*0}(2P)\ell^+\nu_{\ell}$ (middle), and $B^{+}\to \bar{D}_2^{*0}(1F)\ell^+\nu_{\ell}$ (right). }
\label{1P}
\end{figure}

In Fig. \ref{1P}, we provide the differential decay rates $\frac{1}{\Gamma}\frac{d\Gamma}{dx}$ for the $B^{+}\to \bar{D}_2^{*0}(2P)\ell^+\nu_{\ell}$ and $B^{+}\to \bar{D}_2^{*0}(1F)\ell^+\nu_{\ell}$. For comparison, we also present the result of $B^{+}\to \bar{D}_2^{*0}(2460)\ell^+\nu_{\ell}$. Where, the solid line in the middle represents the center value, while the dashed lines on either side indicate the upper and lower limits, respectively. It can be seen that the energy spectrum of the transition $B^{+}\to \bar{D}_2^{*0}(1F)$ is distinctly different from those of the $B^{+}\to \bar{D}_2^{*0}(2460)$ and $B^{+}\to \bar{D}_2^{*0}(2P)$, exhibiting a double-peak structure. Therefore, besides the branching ratios (where the branching ratio for the $2P$ final state is approximately 28 times larger than that for the $1F$ final state), this double-peak structure may help distinguish the $2P$ and $1F$ states experimentally.

The results of nonleptonic decays are provided in Table \ref{III}. Compared to the corresponding semileptonic decay processes, the branching ratios of nonleptonic decays are smaller.  In particular, the values of $\mathcal{B}\left(B\to D_2^{*}(1F)\pi\right)$ are more than an order of magnitude smaller than those of $\mathcal{B}\left(B\to D_2^{*}(1F)\ell^+\nu_{\ell}\right)$.
Thus to search for $D_2^{*}(1F)$ mesons in $B$-meson decays, the semileptonic decay channels are relatively more accessible experimentally than the nonleptonic decay channels.
\begin{table}[!htb]
\caption{Decay widths and branching ratios of nonleptonic decays of $B$ to $2^+$ charmed mesons.}
\begin{tabular}{cccccccc}
\hline
&decay channel&~~~~~&Decay width~(\rm{MeV})&~~~~~&${Br}$& ~~~~~   \\ \hline
&$B^+$ $\rightarrow \bar{D}_2^{*0}(2460)\pi^+$  &&$(2.70^{+0.91}_{-0.67}\pm0.27\pm0.27)\times10^{-13}$&&$(6.73^{+2.27}_{-1.67}\pm0.67\pm0.67)\times10^{-4}$&  \nonumber\\
&${B}^0$ $\rightarrow D_2^{*-}(2460)\pi^+$      &&$(2.78^{+0.91}_{-0.73}\pm0.28\pm0.28)\times10^{-13}$&&$(6.62^{+2.17}_{-1.74}\pm0.66\pm0.66)\times10^{-4}$&  \nonumber\\
&$B^+$ $\rightarrow \bar{D}_2^{*0}(2P)\pi^+$    &&$(1.45^{+0.54}_{-0.42}\pm0.15\pm0.15)\times10^{-14}$&&$(3.62^{+1.34}_{-1.05}\pm0.36\pm0.36)\times10^{-5}$&  \nonumber\\
&${B}^0$ $\rightarrow D_2^{*-}(2P)\pi^+$        &&$(1.33^{+0.76}_{-0.25}\pm0.13\pm0.13)\times10^{-14}$&&$(3.17^{+1.83}_{-0.57}\pm0.32\pm0.32)\times10^{-5}$&   \nonumber\\
&$B^+$ $\rightarrow \bar{D}_2^{*0}(1F)\pi^+$    &&$(7.54^{+0.15}_{-1.57}\pm0.75\pm0.75)\times10^{-17}$&&$(1.88^{+0.04}_{-0.39}\pm0.19\pm0.19)\times10^{-7}$&  \nonumber\\
&${B}^0$ $\rightarrow D_2^{*-}(1F)\pi^+$        &&$(6.67^{+1.02}_{-0.37}\pm0.67\pm0.67)\times10^{-17}$&&$(1.58^{+0.25}_{-0.08}\pm0.16\pm0.16)\times10^{-7}$&  \nonumber\\
&$B^+$ $\rightarrow \bar{D}_2^{*0}(3P)\pi^+$    &&$(3.93^{+1.71}_{-1.09}\pm0.39\pm0.39)\times10^{-16}$&&$(9.80^{+4.30}_{-2.70}\pm0.98\pm0.98)\times10^{-7}$&   \nonumber\\
&${B}^0$ $\rightarrow D_2^{*-}(3P)\pi^+$        &&$(4.70^{+1.13}_{-1.76}\pm0.47\pm0.47)\times10^{-16}$&&$(1.11^{+0.28}_{-0.41}\pm0.11\pm0.11)\times10^{-6}$&   \nonumber\\
&$B^+$ $\rightarrow \bar{D}_2^{*0}(2F)\pi^+$    &&$(2.35^{+2.82}_{-2.18}\pm0.24\pm0.24)\times10^{-16}$&&$(5.86^{+7.04}_{-5.45}\pm0.59\pm0.59)\times10^{-7}$&   \nonumber\\
&${B}^0$ $\rightarrow D_2^{*-}(2F)\pi^+$        &&$(2.70^{+1.93}_{-2.43}\pm0.27\pm0.27)\times10^{-16}$&&$(6.42^{+4.60}_{-5.77}\pm0.64\pm0.64)\times10^{-7}$&   \\
\hline
\end{tabular}\\
\label{III}
\end{table}

\subsection{Partial wave contributions and the mixing}
We note that from the wave function in Eq. (\ref{2+eq}) or Table \ref{2++ partial wave}, both the $nP$ and the $mF$ states ($n=1,2,3$ and $m=1,2$), are $P-D-F$ mixtures. And Table \ref{2++ partial wave} shows that in the $nP$ state, the $P$-wave component predominates, followed by the $D$-wave component, with the $F$-wave being the smallest, for example, $P:D:F=1:0.367:0.063$ for $D_2^{*}(2460)$. In the $mF$ state, the $F$-wave predominates, followed by the $D$-wave, with the $P$-wave being the smallest, for example, $P:D:F=0.074:0.550:1$ for $D_2^{*}(1F)$.

The above information is obtained directly from the wave functions themselves. Next, taking the semileptonic decay as an example, we examine the contributions of each partial wave in the wave functions during the transition. In Tables \ref{B+ to 1P} and \ref{B+ to 1F}, we present the contributions from each partial wave in the initial and final states to the branching ratios of
$B^+ \rightarrow \bar D_2^{*0}(2460)\ell^{+} \nu_{\ell}$ and
$B^+ \rightarrow \bar D_2^{*0}(1F)\ell^{+} \nu_{\ell}$. In the Tables, a prime is added to partial waves of the final state to distinguish them from the initial state. Here, for example, $S\times F'$ means the contribution given solely by the initial $S$-wave and final $F'$-wave, ignoring other partial waves.

\begin{table}[ht]
\centering
\caption{Contributions of partial waves to the branching ratio of $B^+ \rightarrow \bar D_2^{*0}(2460)\ell^{+} \nu_{\ell}$ (in $10^{-3}$).} \label{B+ to 1P}
\setlength{\tabcolsep}{6pt}
\renewcommand{\arraystretch}{1}
\begin{tabular}{|c|c|c|c|c|}
\hline
{\diagbox{$0^-$}{$2^+$}} & whole & $P'$ wave & $D'$ wave ($A_1$,$A_2$,$A_7$,$A_8$) & $F'$ wave \\ \hline
whole                    & 2.67  & 1.22  & 0.334 & 0.0055 \\ \hline
$S$ wave ($B_1$,$B_2$)   & 1.66  & 3.24  & 0.384 & 0.00023 \\ \hline
$P$ wave ($B_3$,$B_4$)   & 0.137  & 0.489  & 1.06 & 0.0042 \\ \hline
\end{tabular}
\end{table}

As can be seen from Table \ref{B+ to 1P}, the non-relativistic contribution $S\times P'$ provides the dominant contribution in $B^+ \rightarrow \bar D_2^{*0}(2460)\ell^{+} \nu_{\ell}$. Its value of 3.24$\times 10^{-3}$ is the largest in Table \ref{B+ to 1P}, indicating that the non-relativistic contribution is larger than the relativistic correction. While the situation is entirely different in Table \ref{B+ to 1F} for the decay $B^+ \rightarrow \bar D_2^{*0}(1F)\ell^{+} \nu_{\ell}$. The relativistic correction from $S\times D'$, with a value of 10.7$\times 10^{-6}$, is the largest contribution, followed by the relativistic correction $P\times F'$ at 7.75$\times 10^{-6}$. Both clearly exceed the non-relativistic term $S\times F'$ contribution of 5.33$\times 10^{-6}$. This indicates that relativistic corrections play a dominant role in this decay channel, consistent with our previous announcement that the higher the excited state, the larger the relativistic correction \cite{19,liwe,wgl1}.

\begin{table}[ht]
\centering
\caption{Contributions of partial waves to the branching ratio of $B^+ \rightarrow \bar D_2^{*0}(1F)\ell^{+} \nu_{\ell}$ (in $10^{-6}$).} \label{B+ to 1F}
\setlength{\tabcolsep}{6pt}
\renewcommand{\arraystretch}{1}
\begin{tabular}{|c|c|c|c|c|}
\hline
{\diagbox{$0^-$}{$2^+$}} & whole & $P'$ wave & $D'$ wave ($A_1$,$A_2$,$A_7$,$A_8$) & $F'$ wave \\ \hline
whole                    & 2.32  & 2.64  & 5.53 & 3.79 \\ \hline
$S$ wave ($B_1$,$B_2$)   & 2.91  & 1.95  & 10.7 & 5.33 \\ \hline
$P$ wave ($B_3$,$B_4$)   & 4.13  & 0.14  & 2.15 & 7.75 \\ \hline
\end{tabular}
\end{table}

In the $1P$ state, the $F$-wave component is minimal ($P:F = 1:0.063$), while in the $1F$ state, the $P'$-wave component is minimal ($P':F' = 0.074:1$). However, these two minimal components exhibit significant differences in transitions. For example, as seen in Table \ref{B+ to 1P}, the contribution 5.5$\times 10^{-6}$ of $(S+P)\times F'$ is the smallest, and much smaller than 1.22$\times 10^{-3}$ from $(S+P)\times P'$, indicating that the $F'$-wave is negligible in the $1P$ state. In contrast, Table \ref{B+ to 1F} shows the contribution of $(S+P)\times P'$ is 2.64$\times 10^{-6}$. Although also the smallest, it is comparable to the $(S+P)\times F'$ contribution of 3.79$\times 10^{-6}$. This demonstrates that the $P$-wave plays a non-negligible role in the $1F$ meson. Therefore, roughly speaking, an $nP$ meson is typically a $P$-$D$ mixture, whereas an $mF$ meson exhibits a $P$-$D$-$F$ mixture due to relativistic correction.
\subsection{Identification of $D^*_2(3000)$ as the $D_2^*(2P)$ meson}
After the new particle $D^*_2(3000)$ is discovered,  numerous theoretical models have studied its mass and full width. However, due to its relatively large mass and abundant decay channels$-$where the primary decay channels involve recently discovered excited states with significant mass uncertainties (e.g., $D^*_1(2600)\pi$ as the dominant decay channel for $D_2^*(2P)$ and $D_2(2740)\pi$ for $D_2^*(1F)$ \cite{txz1})$-$theoretical predictions exhibit substantial uncertainties in the total width. This prevents reliable conclusions regarding $D^*_2(3000)$'s properties, leaving the $D_2^*(2P)$, $D_2^*(1F)$, $D_2^*(2P)$ and $D_2^*(1F)$ assignments as viable candidates.

This uncertain situation can be clarified through the study of $B$ nonleptonic decays.  Using branching ratios $\mathcal{B}\left(\bar{D}_{2}^{*}(2P)^0 \rightarrow D^{-} \pi^{+}\right)=2.52\%$, $\mathcal{B}\left(\bar{D}_{2}^{*}(1F)^0 \rightarrow D^{-} \pi^{+}\right)=1.79\%$, $\mathcal{B}\left(\bar{D}_{2}^{*}(3P)^0 \rightarrow D^{-} \pi^{+}\right)=6.7\%$, and $\mathcal{B}\left(\bar{D}_{2}^{*}(2F)^0 \rightarrow D^{-} \pi^{+}\right)=8.02\%$ calculated in our previous paper \cite{txz1}, we obtain
\begin{eqnarray}
&&\mathcal{B}\left(B^{+} \rightarrow \bar{D}_{2}^{*}(2P)^0 \pi^{+} \right) \mathcal{B}\left(\bar{D}_{2}^{*}(2P)^0\rightarrow D^{-}\pi^{+}\right)=(9.12^{+3.37}_{-2.64}\pm0.91\pm0.91) \times 10^{-7},\nonumber\\
&&\mathcal{B}\left(B^{+} \rightarrow \bar{D}_{2}^{*}(1F)^0 \pi^{+} \right) \mathcal{B}\left(\bar{D}_{2}^{*}(1F)^0\rightarrow D^{-}\pi^{+}\right)=(3.37^{+0.01}_{-0.07}\pm0.34\pm0.34) \times 10^{-9}.\nonumber\\
&&\mathcal{B}\left(B^{+} \rightarrow \bar{D}_{2}^{*}(3P)^0 \pi^{+} \right) \mathcal{B}\left(\bar{D}_{2}^{*}(3P)^0\rightarrow D^{-}\pi^{+}\right)=(6.57^{+0.29}_{-0.18}\pm0.66\pm0.66) \times 10^{-8},\nonumber\\
&&\mathcal{B}\left(B^{+} \rightarrow \bar{D}_{2}^{*}(2F)^0 \pi^{+} \right) \mathcal{B}\left(\bar{D}_{2}^{*}(2F)^0\rightarrow D^{-}\pi^{+}\right)=(4.70^{+5.60}_{-4.40}\pm0.47\pm0.47) \times 10^{-8}.
\end{eqnarray}
The result for the $2P$ state shows excellent agreement with experimental data \cite{2},
$$\mathcal{B}\left(B^{-} \rightarrow \bar{D}_{2}^{*}(3000)^0 \pi^{-} \right) \mathcal{B}\left(\bar{D}_{2}^{*}(3000)^0 \rightarrow D^{+} \pi^{-}\right) =  (2 \pm 3) \times 10^{-6},
$$
while those for the $1F$, $3P$ and $2F$ final states are two or three orders of magnitude smaller than the experimental central value and thus excluded. Therefore, we conclude that the $D^*_2(3000)$ is the $D^*_2(2P)$ meson.

Meanwhile, for the semileptonic decays, we obtain,
\begin{eqnarray}
&&\mathcal{B}\left(B^{+} \rightarrow \bar{D}_{2}^{*}(2P)^0 \ell^{+} \nu_{\ell} \right) \mathcal{B}\left(\bar{D}_{2}^{*}(2P)^0 \rightarrow D^{-} \pi^{+}\right)= (1.76^{+1.10}_{-0.63})  \times 10^{-6}, \nonumber\\
&&\mathcal{B}\left(B^{0} \rightarrow {D}_{2}^{*}(2P)^- \ell^{+} \nu_{\ell} \right) \mathcal{B}\left(D_{2}^{*}(2P)^- \rightarrow  \bar{D}^{0} \pi^{-}\right)   = (1.53^{+1.39}_{-0.38}) \times 10^{-6}.
\end{eqnarray}
Here the branching ratio $\mathcal{B}\left(D_{2}^{*}(2P)\to {D}\pi\right)= 2.53\%$ is also adopted from our previous work \cite{txz1}, which is very close to $2.93\%$ from Ref. \cite{cankaoshu3} and $2.55\%$ from Ref. \cite{lx2}.
This semileptonic decay process is approximately twice as large as the corresponding nonleptonic process, for example, $$\frac{\mathcal{B}\left(B^{+} \rightarrow \bar{D}_{2}^{*}(2P)^0 \ell^{+} \nu_{\ell} \right) }{\mathcal{B}\left(B^{+} \rightarrow \bar{D}_{2}^{*}(2P)^0 \pi^{+} \right) }=1.94^{+0.36}_{-0.19}\pm0.19\pm0.19.$$ Being relatively easier to detect in experiment, it provides a viable decay channel for further investigation of ${D}_{2}^{*}(2P)$ (${D}_{2}^{*}(3000)$) experimentally, with the expectation of reducing the mass and width uncertainties of the ${D}_{2}^{*}(3000)$.
\subsection{Search for the $D^*_2(1F)$ meson}
Now that we have confirmed that $D^*_2(3000)$ is $D^*_2(2P)$, the next question is how to search for the undiscovered mesons $D^*_2(1F)$, $D^*_2(3P)$, and $D^*_2(2F)$ experimentally. For $D^*_2(1F)$, since the branching ratio $\mathcal{B}\left(B^{+} \rightarrow \bar{D}_{2}^{*}(1F)^0 \ell^{+} \nu_{\ell} \right)=(2.32^{+1.69}_{-1.02})\times10^{-6}$ is one order of magnitude larger than $\mathcal{B}\left(B^{+} \rightarrow \bar{D}_{2}^{*}(1F)^0 \pi^{+}\right)=(1.88^{+0.04}_{-0.39}\pm0.19\pm0.19)\times10^{-7}$, the  semileptonic process $B^{+} \rightarrow \bar{D}_{2}^{*}(1F)^0 \ell^{+} \nu_{\ell}$ is a more favorable channel.

For experimental detection convenience, we also need to discuss the issue of the cascade decays of the $D_2^{*}(1F)$ meson. Given that the mass is approximately 3100 MeV (3053$\sim$3132 MeV \cite{3,cankaoshu3,5,6,7,eichten1,ymz1,lcd1,cankaoshu7}), $D_2^{*}(1F)$ has numerous OZI-allowed strong decay channels \cite{9,10,11,12,13,14,15,18}. In our previous paper \cite{txz1}, we calculated the strong decays of $D^*_2(3000)$ as a $2^3P_2$ $3^3P_2$, $1^3F_2$, or $2^3F_2$ meson. We found that the dominant decay channel of $D_2^{*}(1F)$ is $D_2^{*0}(1F)\to D_2(2740)^+\pi^-$, where $D_2(2740)^+$ represents a $1D$ charmed meson, followed by $D_2^{*0}(1F)\to D_1(2420)^+\pi^-$, where $D_1(2420)^+$ corresponds to the $1P$ charmed meson.
However, since $1D$ and $1P$ charmed mesons decay rapidly and are challenging to reconstruct experimentally, the decay channel $D_2^{*}(1F)\to D^+\pi^-$ with a relatively smaller branching ratio may be more experimentally accessible. Therefore, we recommend that future experiments search for $D_2^{*0}(1F)$ in the $B$ semileptonic decay followed by the cascade decay of $D_2^{*0}(1F)\to D^+\pi^-$.
If we choose the branching ratio $\mathcal{B}\left(D_{2}^{*}(1F) \to {D}\pi\right)=1.79\%$ \cite{txz1}, then obtain
\begin{eqnarray}\label{1F}
&&\mathcal{B}\left(B^{+} \rightarrow \bar{D}_{2}^{*}(1F)^0 \ell^{+} \nu_{\ell} \right) \mathcal{B}\left(\bar{D}_{2}^{*}(1F)^0 \rightarrow D^{-} \pi^{+}\right) =  (4.15^{+3.02}_{-1.83}) \times 10^{-8}, \nonumber\\
&&\mathcal{B}\left(B^{0} \rightarrow {D}_{2}^{*}(1F)^- \ell^{+} \nu_{\ell} \right) \mathcal{B}\left(D_{2}^{*}(1F)^- \rightarrow  \bar{D}^{0} \pi^{-}\right)  = (4.03^{+2.08}_{-1.75}) \times 10^{-8}.
\end{eqnarray}
The obtained results are very small, partly because we adopted a relatively small branching ratio for $\bar{D}_{2}^{*}(1F)^0 \rightarrow D^{-} \pi^{+}$. Due to the large width of $D_2^{*}(1F)$, the values of this branching ratio given by different theoretical results also vary significantly; for instance, our result is $1.79\%$ \cite{txz1}, while Ref. \cite{cankaoshu9} gives $3.26\%$, Ref. \cite{cankaoshu7} $3.67\%$, Ref. \cite{cankaoshu3} $6.33\%$, and Ref. \cite{lx2} $7.84\%$, etc. If we choose the value of $7.84\%$, the values in Eq. (\ref{1F}) will become to $1.8 \times 10^{-7}$. Although still small, when considering solely the branching ratio, this process might be detectable in experiment. However, another difficulty arises: theoretical predictions indicate that the $D_2^{*}(1F)$ meson possesses a very large width. For example, with a $D_2^{*}(1F)$ mass around 3100 MeV, we obtained a width of 523 MeV \cite{txz1}, while Ref. \cite{cankaoshu7} gives 722 MeV, Ref. \cite{zxh1} gives 900 MeV, and even the smallest is 243 MeV \cite{cankaoshu3}. Such a large width presents a significant challenge for experimental investigation.

\subsection{Search for the $D^*_2(3P)$ meson}
Regarding the $D^*_2(3P)$ meson, its theoretical mass is approximately 3300 MeV. Its branching ratios in the semileptonic and nonleptonic decays of $B$ mesons are $\mathcal{B}(B^{+}\rightarrow \bar{D}_2^{*0}(3P)\ell^+ \nu_{\ell})=(1.44^{+0.71}_{-0.44})\times10^{-6}$, $\mathcal{B}({B}^{0}\rightarrow D_2^{*-}(3P)\ell^+ \nu_{\ell})=(1.67^{+0.44}_{-0.74})\times10^{-6}$, $\mathcal{B}(B^+\rightarrow \bar{D}_2^{*0}(3P)\pi^+)=(9.80^{+4.30}_{-2.70}\pm0.98\pm0.98)\times10^{-7}$,
and $\mathcal{B}({B}^0\rightarrow D_2^{*-}(3P)\pi^+)=(1.11^{+0.28}_{-0.41}\pm0.11\pm0.11)\times10^{-6}$, with comparable values. In contrast to the very broad state $D_2^{*}(1F)$, $D^*_2(3P)$ is assessed as a relatively narrow state, with its width reported as 116 MeV \cite{cankaoshu3}, 62.57 MeV \cite{cankaoshu9}, and 102.4 MeV \cite{7}, while our estimate yields 35 MeV \cite{txz1}. Its strong decay branching ratio $\mathcal{B}(\bar{D}_{2}^{*}(3P)^0 \rightarrow D^{-} \pi^{+})$ is calculated to be $6.70\%$ \cite{txz1}, therefore we obtain,
\begin{eqnarray}\label{3P}
&&\mathcal{B}\left(B^{+} \rightarrow \bar{D}_{2}^{*}(3P)^0 \ell^{+} \nu_{\ell} \right) \mathcal{B}\left(\bar{D}_{2}^{*}(3P)^0 \rightarrow D^{-} \pi^{+}\right) =(0.96^{+0.48}_{-0.29}) \times 10^{-7}, \nonumber\\
&&\mathcal{B}\left(B^{0} \rightarrow {D}_{2}^{*}(3P)^- \ell^{+} \nu_{\ell} \right) \mathcal{B}\left(D_{2}^{*}(3P)^- \rightarrow  \bar{D}^{0} \pi^{-}\right)  =(1.12^{+0.29}_{-0.50}) \times 10^{-7},\nonumber\\
&&\mathcal{B}\left(B^{+} \rightarrow \bar{D}_{2}^{*}(3P)^0 \pi^{+}  \right) \mathcal{B}\left(\bar{D}_{2}^{*}(3P)^0 \rightarrow D^{-} \pi^{+}\right) =(6.57^{+2.88}_{-1.81}\pm0.66\pm0.66) \times 10^{-8}, \nonumber\\
&&\mathcal{B}\left(B^{0} \rightarrow {D}_{2}^{*}(3P)^- \pi^{+}  \right) \mathcal{B}\left(D_{2}^{*}(3P)^- \rightarrow  \bar{D}^{0} \pi^{-}\right)  = (7.44^{+1.88}_{-2.75}\pm0.74\pm0.74) \times 10^{-8}.
\end{eqnarray}
These values are comparable to that of $\mathcal{B}(B^{+} \rightarrow \bar{D}_{2}^{*}(1F)^0 \ell^{+} \nu_{\ell} ) \mathcal{B}(\bar{D}_{2}^{*}(1F)^0 \rightarrow D^{-} \pi^{+})$. However, considering that the ${D}_{2}^{*}(1F)$ state is broad state while ${D}_{2}^{*}(3P)$ is narrow, the probability of experimentally finding ${D}_{2}^{*}(3P)$ is much higher than that for ${D}_{2}^{*}(1F)$.

\subsection{Search for the $D^*_2(2F)$ meson}
Although the mass of $D^*_2(2F)$ is larger than that of $D^*_2(1F)$, the presence of a node in its wave function results in a significantly smaller total width compared to $D^*_2(1F)$. However, due to its higher mass, about 3350 MeV, and the availability of more decay channels, its width is relatively broad. Theoretical estimates for its width are: 223 MeV \cite{cankaoshu3}, 302.2 MeV \cite{7}, and 193 MeV \cite{txz1}. Further using the branching ratio $\mathcal{B}(\bar{D}_{2}^{*}(2F)^0 \rightarrow D^{-} \pi^{+})=8.02\%$ that we provided \cite{txz1}, we obtain
\begin{eqnarray}\label{2F}
&&\mathcal{B}\left(B^{+} \rightarrow \bar{D}_{2}^{*}(2F)^0 \ell^{+} \nu_{\ell} \right) \mathcal{B}\left(\bar{D}_{2}^{*}(2F)^0 \rightarrow D^{-} \pi^{+}\right) =  (5.23^{+5.35}_{-4.90})\times 10^{-8}, \nonumber\\
&&\mathcal{B}\left(B^{0} \rightarrow {D}_{2}^{*}(2F)^- \ell^{+} \nu_{\ell} \right) \mathcal{B}\left(D_{2}^{*}(2F)^- \rightarrow  \bar{D}^{0} \pi^{-}\right)  = (5.31^{+5.20}_{-4.72}) \times 10^{-8},\nonumber\\
&&\mathcal{B}\left(B^{+} \rightarrow \bar{D}_{2}^{*}(2F)^0 \pi^{+}  \right) \mathcal{B}\left(\bar{D}_{2}^{*}(2F)^0 \rightarrow D^{-} \pi^{+}\right) =  (4.70^{+5.65}_{-4.37}\pm0.47\pm0.47) \times 10^{-8}, \nonumber\\
&&\mathcal{B}\left(B^{0} \rightarrow {D}_{2}^{*}(2F)^- \pi^{+}  \right) \mathcal{B}\left(D_{2}^{*}(2F)^- \rightarrow  \bar{D}^{0} \pi^{-}\right)  = (5.15^{+3.67}_{-4.63}\pm0.52\pm0.52) \times 10^{-8}.
\end{eqnarray}
Although the values of these branching ratios are slightly smaller than those of the corresponding ${D}_{2}^{*}(3P)$ state, the considerably larger total width makes experimental detection significantly more challenging. In summary, the prospects for experimentally discovering the $D^*_2(2F)$ state are less promising than for ${D}_{2}^{*}(3P)$, but easier than for $D^*_2(1F)$.

\subsection{Discussion on mass dependence}
\begin{figure}[!htb]
\begin{minipage}[c]{1\textwidth}
\includegraphics[width=3in]{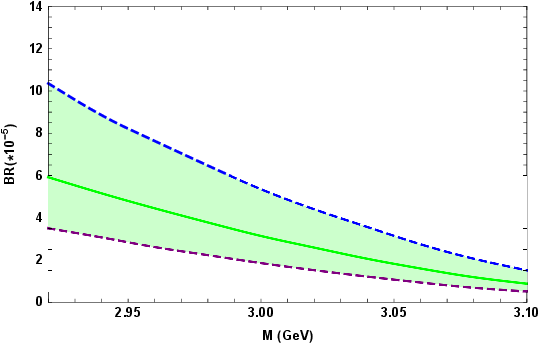}
\includegraphics[width=3in]{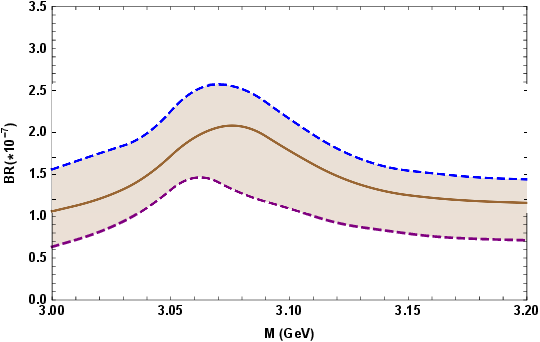}
\end{minipage}
\caption{The nonleptonic decay branching ratios along with the masses of $2^+$ mesons. The left panel is $B^{+}\rightarrow \bar{D}_2^{*0}(2P)\pi^{+}$, the right one is  $B^{+}\rightarrow \bar{D}_2^{*0}(1F)\pi^{+}$.  }
\label{Nwave}
\end{figure}

\begin{figure}[!htb]
\begin{minipage}[c]{1\textwidth}
\includegraphics[width=3in]{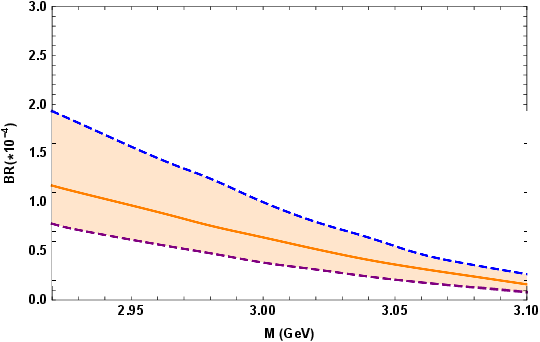}
\includegraphics[width=3in]{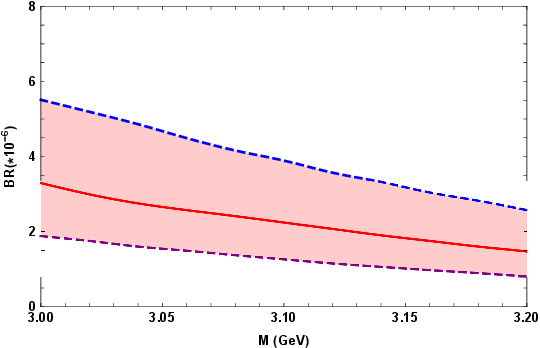}
\end{minipage}
\caption{Semileptonic decay branching ratios along with the masses of $2^+$ mesons. The left panel is $B^{+}\rightarrow \bar{D}_2^{*0}(2P)\ell^{+}\nu_{\ell}$, the right one is $B^{+}\rightarrow \bar{D}_2^{*0}(1F)\ell^{+}\nu_{\ell}. $ }
\label{wave}
\end{figure}

Due to the presence of node, our previous study revealed that theoretical result of $B$ decay to $D_0(2P)$ exhibits a strong dependence on its mass \cite{txz2}.
Similarly, the ${D}_{2}^{*}(2P)$ wave function possesses a nodal structure, see Fig. \ref{2+}, the theoretical decay widths for the processes it participates in will also exhibit a strong dependence on parameters. This can be seen from the significant differences between the decay widths $\Gamma(B^{+} \rightarrow \bar{D}_{2}^{*}(2P)^0 \ell^{+} \nu_{\ell})=(2.81^{+1.78}_{-1.00})\times10^{-14}$ and $\Gamma(B^{0} \rightarrow {D}_{2}^{*}(2P)^- \ell^{+} \nu_{\ell})=(2.55^{+2.31}_{-0.63})\times10^{-14}$ in Table \ref{II}, as well as the significant differences between the $\Gamma(B^{+} \rightarrow \bar{D}_{2}^{*}(2P)^0 \pi^{+})=(1.45^{+0.54}_{-0.42}\pm0.15\pm0.15)\times10^{-14}$ and $\Gamma({B}^0\rightarrow D_2^{*-}(2P)\pi^+)=(1.33^{+0.76}_{-0.25}\pm0.13\pm0.13)\times10^{-14}$ in Table \ref{III}. Given that the experimental mass of the ${D}_{2}^{*}(2P)$ still has large uncertainties, we plot the dependence of the $B$ nonleptonic and semileptonic decay branching ratios on the mass of ${D}_{2}^{*}(2P)$ in Fig. \ref{Nwave} and Fig. \ref{wave}, respectively.

However, we note that despite being a highly excited state with $L=3$, the wave function of $D^*_2(1F)$ is node-free, see Fig. \ref{2+}. The decays of $B$ to $D^*_2(1F)$ are expected to be not very sensitive to mass variations, so
we also include its mass-dependence plots in Fig. \ref{Nwave} and Fig. \ref{wave}.

It can be seen that from the Figures that the results for the ${D}_{2}^{*}(2P)$ exhibit a strong dependence on its mass. When the mass varies from 2.92 GeV to 3.10 GeV, both nonleptonic and semileptonic decay branching ratios decrease by approximately a factor of six. In contrast, while the corresponding processes involving ${D}_{2}^{*}(1F)$ undergo mass variations over a wider range, from 3.0 GeV to 3.20 GeV, the nonleptonic and semileptonic branching ratios vary by less than a factor of two. 

\section{Summary}
In this paper, we employ the Bethe-Salpeter equation approach to calculate the semileptonic and nonleptonic decays of $B$ meson to charmed mesons with quantum numbers $J^P = 2^+$. We first studied the process with the $D^*_2(2460)$ final state, validating the effectiveness of our method through comparison with experimental data. Subsequently, we investigated processes with the ${D}_{2}^{*}(2P)$, ${D}_{2}^{*}(1F)$, ${D}_{2}^{*}(3P)$ and ${D}_{2}^{*}(2F)$ final states. Our study reveals that the discovered particle ${D}_{2}(3000)$, which could not be conclusively identified via strong decays, exhibits good agreement with the predictions for the ${D}_{2}^{*}(2P)$. Other potential candidates, including $D^*_2(1F)$, $D^*_2(3P)$, and $D^*_2(2F)$, are ruled out as their theoretically predicted production branching ratios in $B$ decays are smaller than the experimental value by 2$-$3 orders of magnitude.
We calculated the discovery potential of the undiscovered states $D^*_2(1F)$, $D^*_2(3P)$, and $D^*_2(2F)$ in $B$ meson decays, obtaining branching fractions on the order of $10^{-6}$ to $10^{-7}$. Though small, $D^*_2(3P)$$-$being a narrow width particle$-$could be the first to be observed, whereas the very broad state $D^*_2(1F)$ is the least likely to be resolved.

\vspace{0.7cm} {\bf Acknowledgments}

This work was supported by the National Natural Science Foundation of China (NSFC) under the Grants No. 12575097 and No. 12075073. W. Li is also supported by Hebei Agricultural University introduced talent research special project (No. YJ2024038).

\section{APPENDIX A}
The positive-energy wave function of the $0^-$ meson is expressed as follows,
\begin{eqnarray}
\varphi_{P}^{++}=\left(B_{1}\frac{\slashed{P}}{M}+B_{2}
+B_{3}\frac{\slashed{q}_{_{\bot}}}{M}+B_{4}\frac{\slashed{P}\slashed{q}_{_{\bot}}}{M^2}\right)\gamma_{5},
\end{eqnarray}
where
$$
B_1=\frac{M}{2}(h_1+h_2\frac{m_1+m_2}{\omega_{_{1}}+\omega_{_{2}}}),\quad\quad\quad B_2=\frac{\omega_{_{1}}+\omega_{_{2}}}{m_1+m_2}B_1,$$$$
B_3=\frac{M(m_2-m_1)}{m_1\omega_{_{2}}+m_2\omega_{_{1}}}B_1,\quad\quad\quad B_4=\frac{M(\omega_{_{2}}+\omega_{_{1}})}{m_1\omega_{_{2}}+m_2\omega_{_{1}}}B_1.
$$

The positive energy wave function of $2^+$ state can be written as
\begin{eqnarray}
\varphi_{2^{+}}^{++}(q_{_\bot})=&\varepsilon_{\mu\nu}q_{\bot}^{\mu}q_{\bot}^{\nu}\ \left(A_{1}+A_{2}\frac{\slashed{P}}{M}+A_{3}\frac{\slashed{q}_{\bot}}{M}+
A_{4}\frac{\slashed{P}\slashed{q}_{\bot}}{M^{2}}\right)\nonumber\\&
+M\varepsilon_{\mu\nu}q_{\bot}^{\nu}\gamma^{\mu}
\left(A_{5}+A_{6}\frac{\slashed{P}}{M}+A_{7}\frac{\slashed{q}_{\bot}}{M}+A_{8}\frac{\slashed{P}\slashed{q}_{\bot}}{M^{2}}\right),
\end{eqnarray}
where we have the following expressions
\begin{eqnarray}
&&A_{1}=\frac{1}{2M(m_{1}\omega_{2}+m_{2}\omega_{1})}[(\omega_{1}+\omega_{2})q_{\bot}^{2}f_{3}+(m_{1}+m_{2})q_{\bot}^{2}f_{4}+2M^{2}\omega_{2}f_{5}-2M^{2}m_{2}f_{6}],\nonumber\\
&&A_{2}=\frac{1}{2M(m_{1}\omega_{2}+m_{2}\omega_{1})}[(m_{1}-m_{2})q_{\bot}^{2}f_{3}+(\omega_{1}-\omega_{2})q_{\bot}^{2}f_{4}-2M^{2}m_{2}f_{5}+2M^{2}\omega_{2}f_{6}],\nonumber\\
&&A_{3}=\frac{1}{2}(f_{3}+\frac{m_{1}+m_{2}}{\omega_{2}+\omega_{1}}f_{4}-\frac{2M^{2}}{m_{1}\omega_{2}+m_{2}\omega_{1}}f_{6}),\nonumber\\
&&A_{4}=\frac{1}{2}(f_{3}+\frac{m_{1}+m_{2}}{\omega_{2}+\omega_{1}}f_{4}-\frac{2M^{2}}{m_{1}\omega_{2}+m_{2}\omega_{1}}f_{6}),\nonumber\\
&&A_{5}=\frac{1}{2}(f_{5}-\frac{\omega_{2}+\omega_{1}}{m_{1}+m_{2}}f_{6}),\quad\quad A_{6}=\frac{1}{2}(f_{6}-\frac{{m_{1}+m_{2}}}{\omega_{2}+\omega_{1}}f_{5}),\nonumber\\
&&A_{7}=\frac{M(\omega_{1}-\omega_{2})}{2(m_{2}\omega_{1}+m_{1}\omega_{2})}(f_{5}-\frac{\omega_{2}+\omega_{1}}{m_{1}+m_{2}}f_{6}),\nonumber\\
&&A_{8}=\frac{M(m_{1}-m_{2})}{2(m_{2}\omega_{1}+m_{1}\omega_{2})}(-f_{5}+\frac{\omega_{2}+\omega_{1}}{m_{1}+m_{2}}f_{6}).
\end{eqnarray}

\section{APPENDIX B}
Since the wave function is relativistic, to avoid double counting, the interaction potential must be non-relativistic when solving the Salpeter equation.
We adopt the Cornell potential, which consists of a linear confining potential plus a Coulomb potential arising from single-gluon exchange
\begin{equation}
V(r)= \lambda r+V_0- \gamma_0\otimes\gamma^0\frac{4}{3}\frac{\alpha_s}{r},\label{potential}
\end{equation}
where $\lambda=0.25$ $\rm GeV^2$ is the string tension, $V_0$ is a free constant appearing in potential model to fit data, $\alpha_s$ is the running coupling constant. To avoid divergence in momentum space and account for the screening effect, a factor $e^{-\alpha r}$ ($\alpha=0.06$ $\rm GeV$) is added \cite{Ding,Chao}:
\begin{equation}
V(r)= \frac{\lambda}{\alpha}(1-e^{-\alpha
r})+V_0-\gamma_0\otimes\gamma^0\frac{4}{3}\frac{\alpha_s}{r}e^{-\alpha
r}.\label{potential1}
\end{equation}
And its representation in momentum space is \cite{0-}:
\begin{eqnarray}\label{potential2}
V(\vec{q})=&&V_{s}(\vec{q})+V_{v}(\vec{q})\gamma_{0}\otimes\gamma^{0},
\nonumber\\
V_{s}(\vec{q})=&&-(\frac{\lambda}{\alpha}+V_{0})\delta^{3}(\vec{q})+\frac{\lambda}{\pi^{2}}\frac{1}{(\vec{q}^{2}+\alpha^{2})^{2}},
\nonumber\\
V_{v}(\vec{q})=&&-\frac{2}{3\pi^{2}}\frac{\alpha_{s}(\vec{q})}{(\vec{q}^{2}+\alpha^{2})},
\nonumber\\
\alpha_{s}(\vec{q})=&&\frac{12\pi}{27}\frac{1}{log(a+\frac{\vec{q}^{2}}{\Lambda_{QCD}^{2}})},
\end{eqnarray}
where $\Lambda_{QCD}=0.24$ $\rm GeV$ is confinement energy scale and $a=e=2.7183$.

\end{document}